\documentclass[fleqn,usenatbib]{mnras}

\usepackage{newtxtext,newtxmath}

\usepackage[T1]{fontenc}
\usepackage{ae,aecompl}

\usepackage{graphicx}	
\usepackage{amsmath}	
\usepackage{amssymb}	
\usepackage{wrapfig}
\usepackage[utf8]{inputenc}
\usepackage{makecell}
\usepackage[utf8]{inputenc}
\usepackage[french]{babel}
\usepackage{lscape}
\usepackage[flushleft]{threeparttable}
\usepackage{pdflscape}
\usepackage{caption}
\edef\hc{\string:} 
\edef\scp{\string;} 

\newcommand{\I}{\,{\sevensize I}}
\newcommand{\II}{\,{\sevensize II}}
\newcommand{\III}{\,{\sevensize III}}

\title[SIGNALS: I. Survey Description]{SIGNALS: I. Survey Description}

\author[Rousseau-Nepton et al.]{L. Rousseau-Nepton$^{1,2}$\thanks{E-mail: SIGNALS-CoPI@cfht.hawaii.edu}, R. P. Martin$^{2}$, 
C. Robert$^{3,2,1}$, 
L. Drissen$^{3,2,1}$, 
P. Amram$^{4}$,  
\newauthor S. Prunet$^{1}$, T. Martin$^{3}$, I. Moumen$^{1,3}$, A. Adamo$^{5}$, A. Alarie$^{3,6}$, P. Barmby$^{7}$,  
\newauthor A. Boselli$^{4}$, F. Bresolin$^{8}$, M. Bureau$^{9}$, L. Chemin$^{10}$, R. C. Fernandes$^{11}$, 
\newauthor F. Combes$^{12,13}$, C. Crowder$^{1}$, L. Della Bruna$^{5}$, F. Egusa$^{14}$, B. Epinat$^{4}$, 
\newauthor V. F. Ksoll$^{15}$, M. Girard$^{16}$, V. G\'omez Llanos$^{17}$
D. Gouliermis$^{15,18}$, 
\newauthor K. Grasha$^{19,20}$, C. Higgs$^{21,22}$, J. Hlavacek-Larrondo$^{23}$, I.-T. Ho$^{18}$, 
\newauthor J. Iglesias-P\'aramo$^{24,25}$, G. Joncas$^{3}$, Z. S. Kam$^{26,27}$, P. Karera$^{3}$, R. C. Kennicutt$^{28}$, 
\newauthor R. S. Klessen$^{15}$, S. Lianou$^{7,29}$, L. Liu$^{9}$, Q. Liu$^{30}$, A. Luiz de Amorim$^{11}$,  
\newauthor J. D. Lyman$^{31}$, H. Martel$^{3}$, B. Mazzilli-Ciraulo$^{12,13}$,  A. F. McLeod$^{32,33}$, 
\newauthor A-L. Melchior$^{12}$, I. Millan$^{34}$, M. Moll\'a$^{34}$, R. Momose$^{35,36}$, C. Morisset$^{17}$, 
\newauthor H.-A. Pan$^{37}$, A. K. Pati$^{38}$, A. Pellerin$^{39}$, E. Pellegrini$^{15}$, I. P\'erez$^{40,41}$, 
\newauthor A. Petric$^{1,42}$, H. Plana$^{43}$, D. Rahner$^{15}$, T. Ruiz Lara$^{44,45}$, L. S\'anchez-Menguiano$^{44}$, 
\newauthor 
K. Spekkens$^{46}$, G. Stasi\'nska$^{47}$, M. Takamiya$^{2}$, N. Vale Asari$^{11,48,49}$, 
\newauthor and J. M. V\'ilchez$^{23}$.
\\
\\
\footnotesize{Affiliations can be found after the references}
}

\date{Accepted XXX. Received YYY; in original form ZZZ}

\pubyear{2019}

\begin{document}
\label{firstpage}
\pagerange{\pageref{firstpage}--\pageref{lastpage}}
\maketitle

\begin{abstract}

SIGNALS, the {\bf S}tar formation, {\bf I}onized {\bf G}as, and {\bf N}ebular {\bf A}bundances {\bf L}egacy {\bf S}urvey, is a large observing program designed to investigate massive star formation and H{\II} regions in a sample of local extended galaxies. The program will use the imaging Fourier transform spectrograph SITELLE at the Canada-France-Hawaii Telescope. Over 355 hours (54.7 nights) have been allocated beginning in fall 2018 for eight consecutive semesters. Once completed, SIGNALS will provide a statistically reliable laboratory to investigate massive star formation, including over 50\,000 resolved H{\II} regions: the largest, most complete, and homogeneous database of spectroscopically and spatially resolved extragalactic H{\II} regions ever assembled. For each field observed, three datacubes covering the spectral bands of the filters SN1 (363\,-\,386\,nm), SN2 (482\,-\,513\,nm), and SN3 (647\,-\,685\,nm) are gathered. The spectral resolution selected for each spectral band is 1000, 1000, and 5000, respectively. As defined, the project sample will facilitate the study of small-scale nebular physics and many other phenomena linked to star formation at a mean spatial resolution of $\sim$20\,pc. This survey also has considerable legacy value for additional topics including planetary nebulae, diffuse ionized gas, and supernova remnants. The purpose of this paper is to present a general outlook of the survey, notably the observing strategy, galaxy sample, and science requirements.
\normalsize

\end{abstract}



\section{Introduction}

Over the past few decades, the astronomical community has expressed a need for integral field spectroscopy. Multiple instruments (i.e. Integral Field Spectrographs; IFS, Imaging Fourier Transform Spectrograph; IFTS, and Fabry-Perot) have been implemented for this purpose in the visible/near-infrared, most notably the Spectroscopic Areal Unit for Research on Optical Nebulae (\texttt{SAURON}, \citealt{Bacon2001}), Fabry-Perot instruments at the Observatoire de Haute-Provence (\citealt{Garrido2002}) and the Observatoire du Mont-M\'egantic (\citealt{Hernandez2003}), the Infrared Spectrograph on the \textit{Spitzer} Space Telescope (IRS, \citealt{Houck2004}), the Potsdam Multi-Aperture Spectrophotometer - fiber Package (PMAS-PPAK, \citealt{Roth2005}), the Visible Integral-Field Replicable Unit Spectrographs prototype (VIRUS-P, \citealt{Hill2008}), the Multi Unit Spectroscopic Explorer (MUSE, \citealt{Laurent2006,Bacon2010}), the Sydney-AAO Multi-object Integral Field Spectrograph (SAMI, \citealt{Bryant2012}), the Sloan Digital Sky Survey IV (SDSS-IV, \citealt{Drory2015}), and the Spectro-Imageur \`a Transform\'ee de Fourier pour l'\'Etude en Long et en Large des raies d'\'Emission (SITELLE, \citealt{Brousseau2014, Drissen2019}). Many other instruments with integral-field capabilities are currently being developed, e.g. the Sloan Digital Sky Survey V (SDSS-V, \citealt{Kollmeier2017}). Several galaxy surveys have been conducted using these instruments while mainly focusing on one specific goal\hc\,\,acquiring a deeper understanding of galaxy evolution from detailed studies of the physical and dynamical properties of their components. Examples include \href{http://www-astro.physics.ox.ac.uk/atlas3d/}{ATLAS$^{3D}$}, Spitzer Infrared Nearby Galaxies Survey (SINGS, \citealt{Daigle2006}), Virgo high-resolution H$\alpha$ kinematical survey (VIRGO, \citealt{chemin2006}), Gassendi H$\alpha$ survey of SPirals (GHASP, \citealt{Garrido2002,epinat2008}), Calar Alto Legacy Integral Field Area survey (\href{http://califa.caha.es/}{CALIFA}, \citealt{Sanchez2012}), VIRUS-P Exploration of Nearby Galaxies (VENGA, \citealt{Blanc2013}), Mapping Nearby Galaxies at Apache Point Observatory (\href{https://www.sdss.org/surveys/manga/}{MANGA}, \citealt{Bundy2015, Yan2016}), 
\href{https://sami-survey.org/}{SAMI survey} \citep{Konstantopoulos2013}, Las Campanas Observatory PrISM Survey (a.k.a. TYPHOON, \citealt{Poetrodjojo2019}), and Physics at High Angular resolution in Nearby Galaxies (\href{https://sites.google.com/view/phangs/home}{PHANGS}, \citealt{Rosolowsky2019}). In these surveys, star formation, chemical enrichment processes, and the gas and stellar kinematics were targeted. These are crucial ingredients in the interaction between stellar populations and the interstellar medium (ISM), and in driving feedback acting as a self-regulatory mechanism on the evolution of galaxies.

These IFS studies made significant progress in our understanding of star-forming galaxies. But, at a spatial resolution ranging from a parsec to several kiloparsecs, these studies had a limited impact on our knowledge of small-scale physics within galaxies. At higher spatial resolutions, the MUSE Atlas of Disks (\href{https://www.phys.ethz.ch/carollo/research/mad.html}{MAD}, \citealt{Erroz-Ferrer2019}) is observing local disk galaxies by focusing on their central parts, providing a good sampling of H{\II} regions in high stellar density environments (center and/or bulge). Additionally, some extensive studies on nearby H{\II} regions (e.g. \citealt{Sanchez2007, McLeod2015, Weilbacher2015}) have paved the way toward a better understanding of the interaction between stars and the different ISM gas phases. Without being exhaustive, we can mention works using mid-infrared and radio observations of the Milky Way (e.g. \citealt{Tremblin2014}) and the H{\II} Region Discovery Survey (\href{https://www.cv.nrao.edu/hrds/}{HRDS},
\citealt{Anderson2011}), 
narrow-band imaging in the Visible of the LMC 
(e.g. Magellanic Cloud Emission Line Survey; \href{http://www.ctio.noao.edu/mcels/}{MCELS}, \citealt{Smith2000, Pellegrini2012}) and of M31 (e.g. A New Catalog of H{\II} Regions in M31, \citealt{Azimlu2011}), or studies using spectral information with IFUs (e.g. with MUSE on NGC\,300, \citealt{Roth2018}). Additionally, the upcoming Local Volume Mapper (\href{https://www.sdss.org/future/lvm/}{LVM}) survey will use SDSS-V to cover the Milky Way and other nearby galaxies. 

Despite these advancements, the need for spectroscopic surveys with a high spatial resolution on a large number of H{\II} regions remains. We need to study a detailed and uniform H{\II} sample in all galactic environments. CFHT's IFTS SITELLE was intentionally designed for this purpose and is currently the most efficient instrument to conduct such a study. With its large 11$^{\prime}$\,$\times$\,11$^{\prime}$ field-of-view (FOV), SITELLE's spatial coverage is 100 times bigger than any current competitor. Compared to fiber-fed systems, SITELLE gathers a much larger fraction of the light coming into the instrument with much better spatial sampling (0.32$^{\prime\prime}$ per pixel) and improved blue sensitivity.

\begin{figure*}
\includegraphics[width=6.8in]{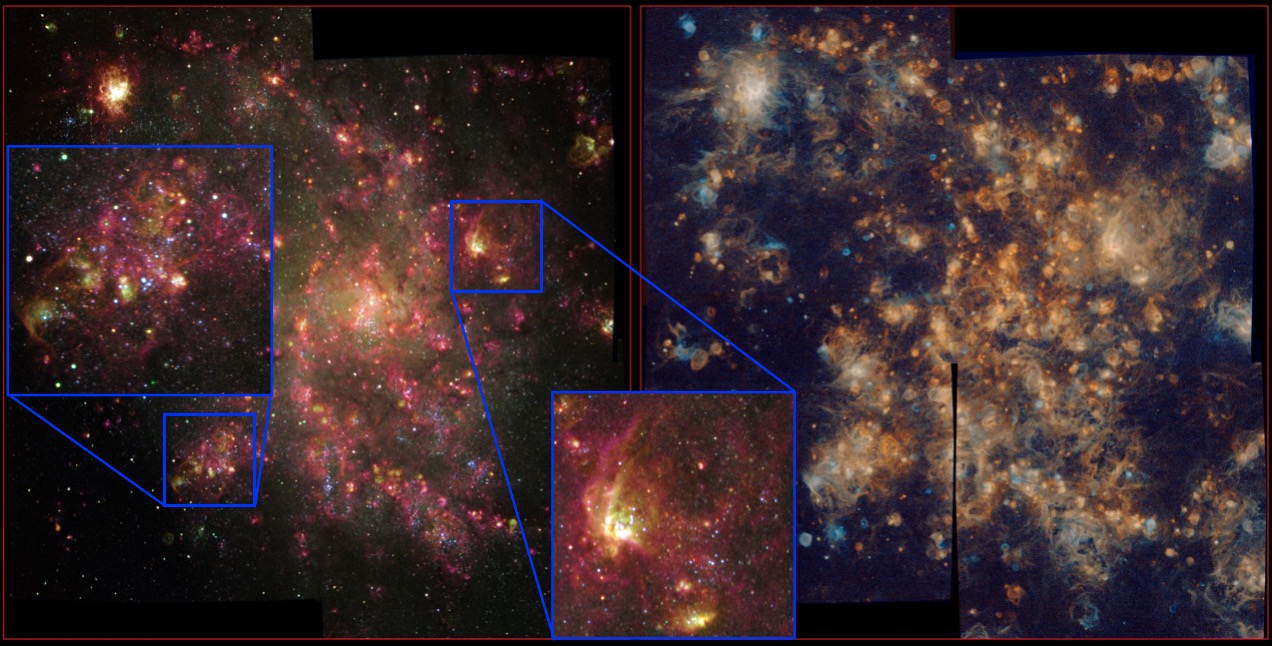} 
\caption{M33: SIGNALS' science verification data (a mosaic of four fields). 
On the left, a composite image using the three filters . It combines the three continua over the wavelength range of the filters plus one emission line for each\hc\,\,SN3+H$\alpha$ (red), SN2+[O\III]$\lambda$5007 (green), and SN1+[O\II]$\lambda$3727 (blue). The blue squares are zooms in two areas to show additional details. On the right, to emphasize the ionized gas component, we used the emission line map of H$\alpha$, [O\III]$\lambda$5007, and [O\II]$\lambda$3727 with different shades of orange, green, and blue, respectively.}
\label{M33}
\end{figure*}

SIGNALS, the {\bf S}tar formation, {\bf I}onized {\bf G}as, and {\bf N}ebular {\bf A}bundances {\bf L}egacy {\bf S}urvey, is using SITELLE to observe more than 50\,000 resolved extragalactic H{\II} regions in a sample of $\sim$40 nearby galaxies over three spectral ranges (363\,-\,386\,nm, 482\,-\,513\,nm, and 647\,-\,685\,nm). By studying the spatially-resolved spectra of individual H{\II} regions and their massive stars content, SIGNALS' main scientific objective focuses on how diverse local environments (nearby stellar population mass and age, dynamic structures, gas density and chemical composition, etc.) affect the star formation process. The legacy of this survey will also contribute to many topics in astrophysics, from stars to very distant galaxies. More specifically, the primary goals of SIGNALS are\hc\,\,

\begingroup{
\leftskip=1.truecm
\parindent=-0.4truecm

1)~to quantify the impact of the surrounding environment on the star formation process\scp

2)~to link the feedback processes to the small-scale chemical enrichment and dynamics in the surrounding of star-forming regions\scp\,\,and 

3)~to measure variations in the resolved star formation rate with respect to indicators used to characterize high-redshift galaxies. 

}\endgroup

\vskip 0.2truecm
This paper presents an overview of the SIGNALS project observational strategy and data reduction (Section~2), galaxy sample (Section~3), science goal requirements (Section~4), data products (Section~5), and legacy value (Section~6). 

\section{Observations and Data Reduction}

\subsection{Strategy and Filter Configurations}
\label{str_filter}

\begin{figure*}
\begin{center}
\includegraphics[trim={0 0.1cm 0 0}, clip, width=6.9in]{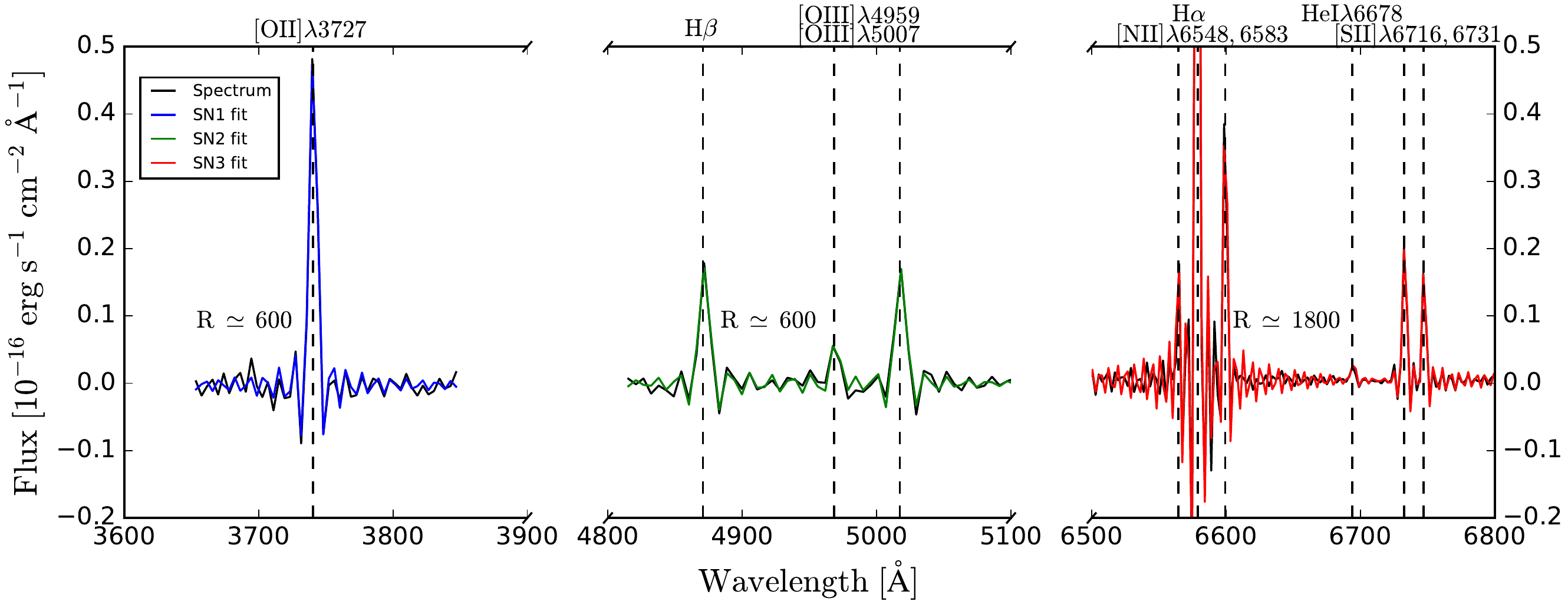} 
\caption{Continuum subtracted spectrum of an H{\II} region in NGC\,628 extracted from the datacubes using a circular aperture with a radius of 1.6$^{\prime\prime}$, centered at RA 01h36m53.1s and DEC +15$^\circ$48$^{\prime}$04.8$^{\prime\prime}$. The fits obtained with ORCS (Sect.~5) for the emission lines are shown. Note that SITELLE's line profiles are fitted using a cardinal sine function. Spectral resolutions obtained here are significantly smaller than those selected for SIGNALS.}
\vspace{-0.2cm}
\label{spectrum}
\end{center}
\end{figure*}

SITELLE was designed to be optimized for the study of emission-line objects. A lot of technical challenges were overcome while building the instrument to assure excellent sensitivity, a high spectral resolution, and a large FOV (\citealt{Drissen2019}). SIGNALS' observing strategy is built upon SITELLE's unique capabilities. Three filters are used to optimize the detection and characterization of the strong diagnostic lines for nearby extragalactic nebulae\hc\,\,SN1 (363\,-\,386\,nm) with the [O\II]$\lambda$3727 and the Balmer emission lines H9 to H12 (detected in bright regions), SN2 (482\,-\,513\,nm) with H$\beta$ and [O\III]$\lambda\lambda$4959,5007, and SN3 (647\,-\,685\,nm) with H$\alpha$, [N\II]$\lambda\lambda$6548,6583, He\I$\lambda$6678, and [S\II]$\lambda\lambda$6717,6731. These filters also provide a sampling of the continuum. Figure~\ref{M33} shows an example of the deep images obtained combining the data from the three SIGNALS' filters using two different methods for M33. This dataset was acquired for the science verification of the SIGNALS project and includes four fields. Figure~\ref{spectrum} shows an example of an H{\II} region spectrum in NGC628 obtained with SITELLE during its commissioning (\citealt{Rousseau-Nepton2018}). The main lines investigated by the survey are compiled in Table~\ref{liner}. The SITELLE Exposure Time Calculator (ETC) was used to simulate the expected signal-to-noise ratio (SNR) for specific lines and to define the instrumental configurations (i.e. spectral resolution, exposure time, and Moon maximal contribution). The ETC SNR calculation includes both the contributions of the sky and the target emission to the photon noise. The main driver for the configurations' selection was the detection threshold (while avoiding saturation of the detector) over the range in surface brightness (SB) observed in H{\II} regions, from $\sim$ 8 $\times$ 10$^{-17}$ to 8 $\times$ 10$^{-12}$\,erg\,s$^{-1}$\,cm$^{-2}$\,arcsec$^{-2}$. Table\,\ref{conf} summarizes the parameters for the three selected configurations. 

\begin{table}
\caption{Main emission lines for SIGNALS}\label{liner}
\centering
\small
\begin{tabular}{|c|c|}
\hline
\small Ion & \small Description \\
\hline
\small [O\II] & \small [O\II]$\lambda\lambda$3726,3729 \\
\hline
\small H\II & \small H$\beta$ (4861\,\AA) \\
\hline
\small [O\III] & \small [O\III]$\lambda$4959 \\
\hline
\small [O\III] & \small [O\III]$\lambda$5007 \\
\hline
\small [N\II] & \small [N\II]$\lambda$6548 \\
\hline
\small H\II & \small H$\alpha$ (6563\,\AA) \\
\hline
\small [N\II] & \small [N\II]$\lambda$6583 \\
\hline
\small He\I & \small He\I$\lambda$6678 \\
\hline
\small [S\II] & \small [S\II]$\lambda$6716 \\
\hline
\small [S\II] & \small [S\II]$\lambda$6731 \\
\hline
\end{tabular}
\end{table}

\begin{table}
\caption{Filter configurations}\label{conf}
\centering
\small
\begin{tabular}{|c|c|c|c|c|c|}
\hline
\small Filter & 
\small \makecell{Band \\
\small [nm]}
& 
\small \makecell{Spectral \\ Res.} & 
\small \makecell{Exposure \\
time \\
\small [sec step$^{-1}$]}
& 
\small \makecell{\# \\
of steps} 
& \small \makecell{Integration \\
time \\
\small [hours]} \\
\hline
\small SN1 & \small 363-386 & \small 1000 & \small 59.0 & \small 172 & \small 3 \\
\hline
\small SN2 & \small 482-513 & \small 1000 & \small 45.5 & \small 219 & \small 3 \\
\hline
\small SN3 & \small 647-685 & \small 5000 & \small 13.3 & \small 842 & \small 4 \\
\hline
\end{tabular}
\end{table}

\begin{figure}
\centering
\includegraphics[width=0.98\linewidth]{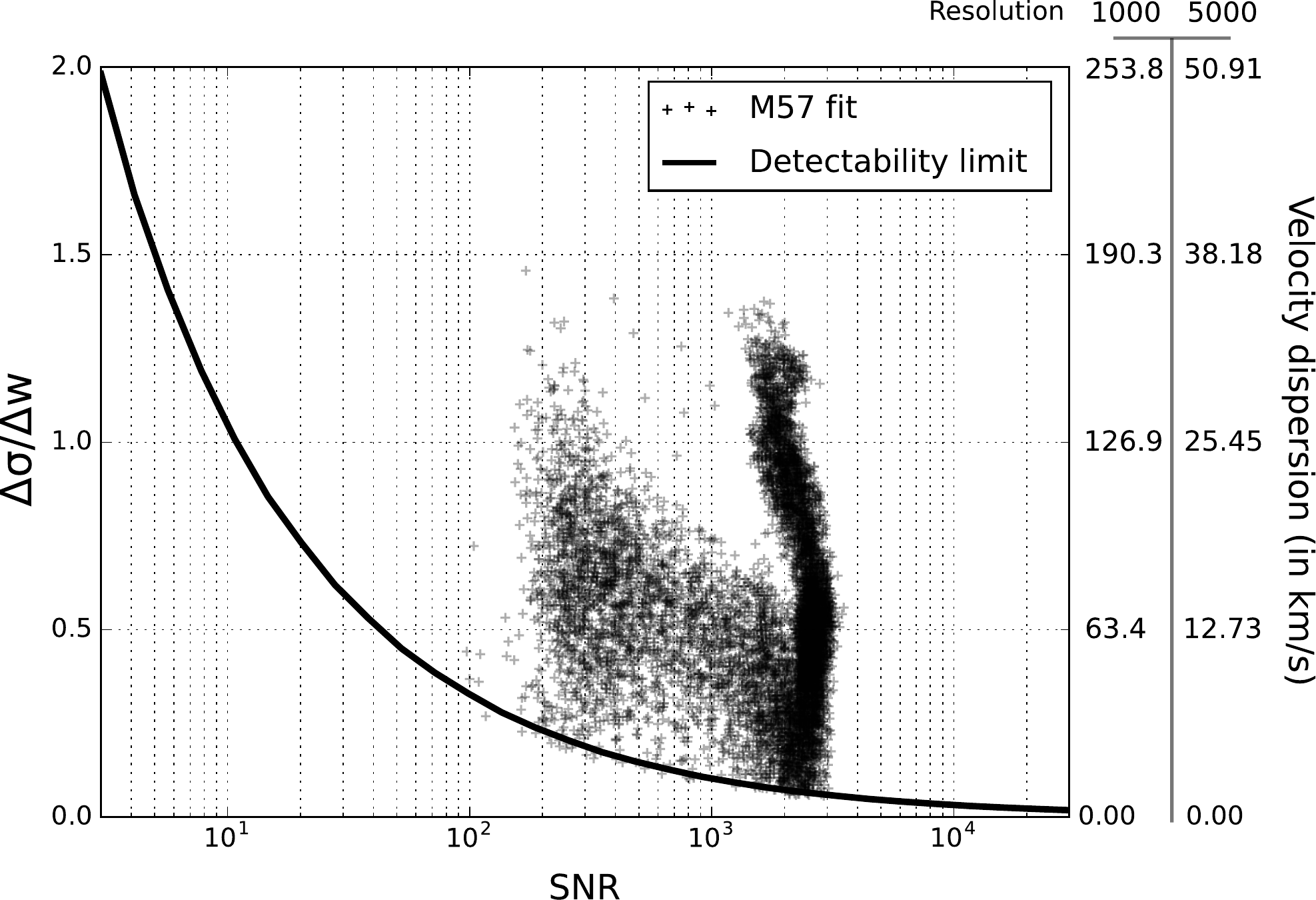}
\caption{Instrument line spread function ratio. The black line shows the minimum measurable ratio $\Delta\sigma$ (observed) /$\Delta w$ (theoretical or pure sinc) for a line with a given SNR. The corresponding velocity dispersion at the two different resolutions used for SIGNALS (5000 and 1000) are reported on the right axis. It scales linearly with the resolution. A set of points from a fit of NGC6720 data with ORCS (\citealt{Martin2017}) is overlapped to the prediction. In this case, we can see that the SNR of the data points is high because all the lines in each spectrum are fitted with the same broadening parameter.}
\label{vel_snr}
\end{figure}

The instrumental configuration for SN1 and SN2 was set using a spectral resolution of 1000, enough to separate the emission lines (the unresolved [OII]$\lambda\lambda$3726,3729 doublet; hereafter [OII]$\lambda$3727, [OIII]$\lambda$4959, [OIII]$\lambda$5007, and H$\beta$) and to properly measure the stellar continua. For the SN3 filter (648\,-\,685\,nm), which contains the H$\alpha$ line broadly used to study gas dynamics, we adopted a higher spectral resolution of R\,=\,5\,000 in order to reach a precision on the velocity measurements of $\sim$0.1 to 10\,km\,s$^{-1}$ (for H{\II} regions with a SNR of 1000 to 10). With this configuration, we are also sensitive to velocity dispersion ranging from 30\,km\,s$^{-1}$ for a faint region (over one seeing element) down to $\sim$\,10 km\,s$^{-1}$ for most of the regions observed and $\sim$\,1-2\,km\,s$^{-1}$ for the brightest resolved regions. Figure\,\ref{vel_snr} shows the relation between the minimum measurable broadening ratio of the Gaussian over the sinc functions used to fit a line of a given SNR. Appropriate binning can be used to reach a higher precision on the velocity measurements for kinematics studies (see Section\,\ref{sec:small-scale-dynamics}). 

With these configurations, the detection thresholds (SNR\,$>\,$3) over an element of resolution of 1$^{\prime\prime}$ are SB$_{\rm H\alpha}$\,$\simeq$\,3.6$\times$10$^{-17}$, 
SB$_{\rm H\beta}$\,$\simeq$\,4.2$\times$10$^{-17}$, 
and SB$_{\rm [O\II]3727}$\,$\simeq$\,3.0$\times$10$^{-17}$\,erg\,s$^{-1}$\,cm$^{-2}$\,arcsec$^{-2}$ for SN3, SN2, and SN1, respectively.
We used H$\alpha$, H$\beta$, and [O\II]$\lambda$3727 as reference since they are the key lines of interest in their respective filters. Note that the threshold for H$\alpha$ corresponds to a SNR of 6.7 on the faintest H{\II} regions and is also good for further investigation of the diffuse ionized gas (DIG, see Section\,\,\ref{feedback}) component at SB$_{\rm H\alpha DIG}$\,$\simeq$\,5$\times$10$^{-18}$\,erg\,s$^{-1}$\,cm$^{-2}$\,arcsec$^{-2}$; additional binning may also be used later during the analysis. No extinction was applied on the ETC simulations but, adopting these numbers, we estimated that over 30\% of the pixels detected in H$\alpha$ will have a H$\beta$ detection before applying any spatial binning. This is also supported by the relative detection we obtained on other galaxies during commissioning of SITELLE. After spatial binning, this detection threshold ensures that most H{\II} regions will have a reliable measurement of the H$\alpha$/H$\beta$ ratio on the integrated spectra ($>$95\%\scp\,\,considering a mean extinction $E(B-V)$\,$\simeq$\,0.5 and no background). 

The current average seeing of the fields that have already been observed is 1.0$^{\prime\prime}$. The seeing of the observations is limited to a mean value of 1.2$^{\prime\prime}$ over a scan in order to maintain the global spatial sampling of the H{\II} regions. Data observed with a mean seeing greater than 1.2$^{\prime\prime}$ are not validated. The moon contribution must be minimal for the blue SN1 filter. For the other two filters the moon contribution can be moderate (50\% illumination and at a distance greater than 70$^{\circ}$) without affecting the detection threshold significantly. 

\subsection{Data Reduction and Calibrations}
\label{datared}

Two software components have been developed specifically for the reduction and spectral analysis of SITELLE's datacubes\hc\,\,the data reduction software, \href{https://github.com/thomasorb/orbs}{ORBS}  and the data analysis software, \href{https://orcs.readthedocs.io/en/latest/}{ORCS} (\citealt{Martin2015,Martin2016}). 

ORBS is a fully automated data reduction pipeline, tailored for IFTS data, transforming the interferogram cube of SITELLE's cameras into a spectral datacube. The first ORBS data processing step includes the wavelength calibration using a He-Ne laser datacube observed during the same observing run as the science datacube. As they come out of the reduction pipeline, datacubes have a pixel-to-pixel velocity error estimated to be of the order of 0.1\,km\,s$^{-1}$ at R\,$=$\,5000. Additional correction is applied to the absolute wavelength calibration using the skylines centroid map as described in \cite{Martin2018} resulting in an accuracy of $\sim$1\,km\,s$^{-1}$ on the measured relative velocities across the FOV (see appendix B of \citealt{Martin2016} for details). 

The first data processing also includes a correction of the zero, first, and greater than one phase orders as described in \cite{Martin2017}. This ensures an instrument line spread (ILS) function that follows the theoretical model according to the selected instrument configuration. The ILS function for SITELLE is a pure cardinal sine (sinc) function. The FWHM of the sinc function is known and fixed since it depends only on the sampling and the maximum optical path displacement of the interferogram scan (\citealt{Martin2016, Drissen2019}). As shown in the work of \cite{Martin2016}, when natural line broadening is observed, a sinc convolved with a Gaussian function can be used to properly recover the line parameters. If so, the line model has three varying parameters (amplitude, velocity, and broadening of the Gaussian). The uncertainty on the measurements depends on the SNR and the proper selection of the line model (sinc or sinc+Gaussian) and the set of lines to be fitted for each pixel (all lines must be fitted simultaneously, and if multiple components are present, they must be fitted simultaneously as well). As also described in \cite{Martin2018}, the precision on the velocity obtained over the planetary nebulae in M31 ranged from 2 to 6\,km\,s$^{-1}$ for sources of surface brightness much lower than what is expected for most H{\II} regions. 

\begin{figure*}
\centering
\includegraphics[trim={0.1cm 0.1cm 0.1cm 0.1cm}, clip, width=0.6\linewidth]{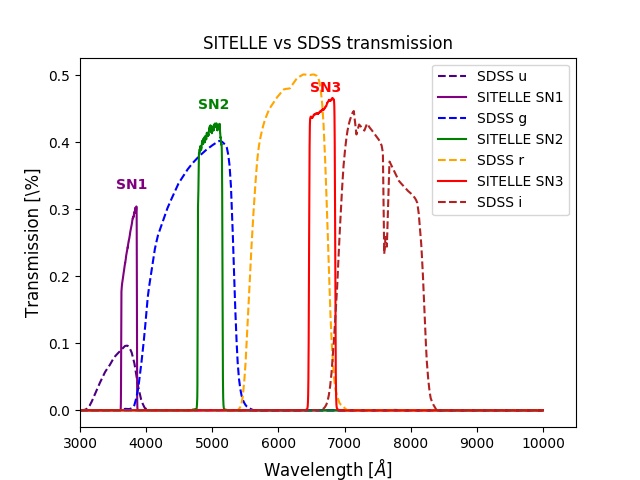}
\caption{Transmittance curves for the atmospheric (airmass=1), telescope and optical assembly from SITELLE and SDSS. The SIGNALS filters (SN1, SN2, and SN3) are shown along with the overlapping SDSS filters.}
\label{SDSS2}
\end{figure*}

The ORBS flux calibration uses flat images and standard star images taken during the same night as the science observation as well as standard star datacubes from the same observation run. The airmass is corrected using the atmospheric extinction curve above Maunakea provided by \cite{Buton2013} and the mean airmass of the cube. Atmospheric extinction variations through a datacube are corrected using the transmission function extracted by normalizing the combined interferogram of the two output detectors to the maximum flux observed through the scan. Since a datacube is observed over a period of 3 to 4 hours and the transmission variations are corrected and normalized to the best transmission conditions seen during this period, the transmission variation is completely negligible. The flux calibration function itself is extracted from the standard star datacube using the star integrated spectrum in a given filter corrected for its own airmass and is used as a reference for the zero point correction. Zero point variations are tracked and corrected using the standard images taken during the night in photometric conditions. This data reduction procedure ensures a pixel-to-pixel relative flux calibration accuracy better than 2\%. The datacube-to-datacube relative flux calibration accuracy is better than 5\% (\citealt{Martin2017}), including both random and systematic errors. This will be improved in the second release of ORBS currently being developed for the SIGNALS program. Figure \ref{SDSS2} shows the relative overlap between the SITELLE and the SDSS filters. Comparison with ancillary data can always be made to improve the flux calibration before the second data release. 

\section{Sample of H{\,\sc ii} Regions}

Our sample selection was driven by the need to observe a very large number of extragalactic H{\II} regions in as many different galactic environments as possible. Star-forming areas within galaxies therefore motivated our choice. They were identified from existing H$\alpha$ and UV-GALEX images from all nearby galaxies observable by CFHT. 
To allow for a minimum contiguous observing period of 2\,h from Maunakea with an airmass of 1.5 or smaller, targets with a declination below $-$22$^\circ$ and over $+$62$^\circ$ (i.e. the limits of the instrument) were rejected. The maximum galaxy distance was set to 10\,Mpc in order to optimize the spatial resolution while still sampling a large number of H{\II} regions. This ensures a spatial resolution of 40\,pc per resolution element or less for a seeing of 0.8$^{\prime\prime}$. Our selection criteria can be summarized as$\hc$

\begingroup
\leftskip=1.truecm
\parindent=-0.4truecm

{1)~star-forming galaxies\scp

2)~$-$22$^\circ$ $<$ DEC $<$ $+$62$^\circ$\scp

3)~D $\leq$ 10\,Mpc\scp

4)~limited amount of dust on the line-of-sight\scp~and

5)~limited crowding of the H{\II} regions (inclination $\leq$ 71$^\circ$).}

\endgroup
\noindent{Here crowding refers to observing multiple HII regions along the line-of-sight. }

Based on these criteria, our sample of potential targets includes 54 objects. Except for some very large objects, most targets necessitate only one SITELLE field. As a result of the first criterion, some non-active areas and entire galaxies were rejected. For instance, galaxies such as the Cetus Dwarf Spheroidal and DDO\,216, as well as the very outskirts of some extended galaxies (like M33) were excluded if they contained fewer than 10 regions within one pointing in order to increase our observing efficiency. To minimize the effect of internal dust extinction, we excluded massive, dusty, edge-on spirals from the sample (e.g. NGC\,4631, NGC\,4244, NGC\,4605, etc.). However, we kept the foreground portion of the M31 disk. Table\,\,\ref{target_list} contains the list of potential targets for the survey and a summary of their properties. Among this sample, 36 fields will be selected and observed with the SITELLE filters SN1, SN2, and SN3, for a total of 355 hours. By extrapolating from available H$\alpha$ images of the galaxies, the number of fields observed, and the known accuracy of the region identification as a function of spatial sampling, we estimate that about 50\,000 H{\II} regions will be analyzed.  Each semester, an updated list will be published on the SIGNALS \href{http://www.signal-survey.org}{Website}\footnote{http://www.signal-survey.org}. 

\begin{figure}
\centering
\vspace{-0.2cm}
\includegraphics[trim={2cm 0.7cm 3cm 1cm}, clip, width=0.98\linewidth]{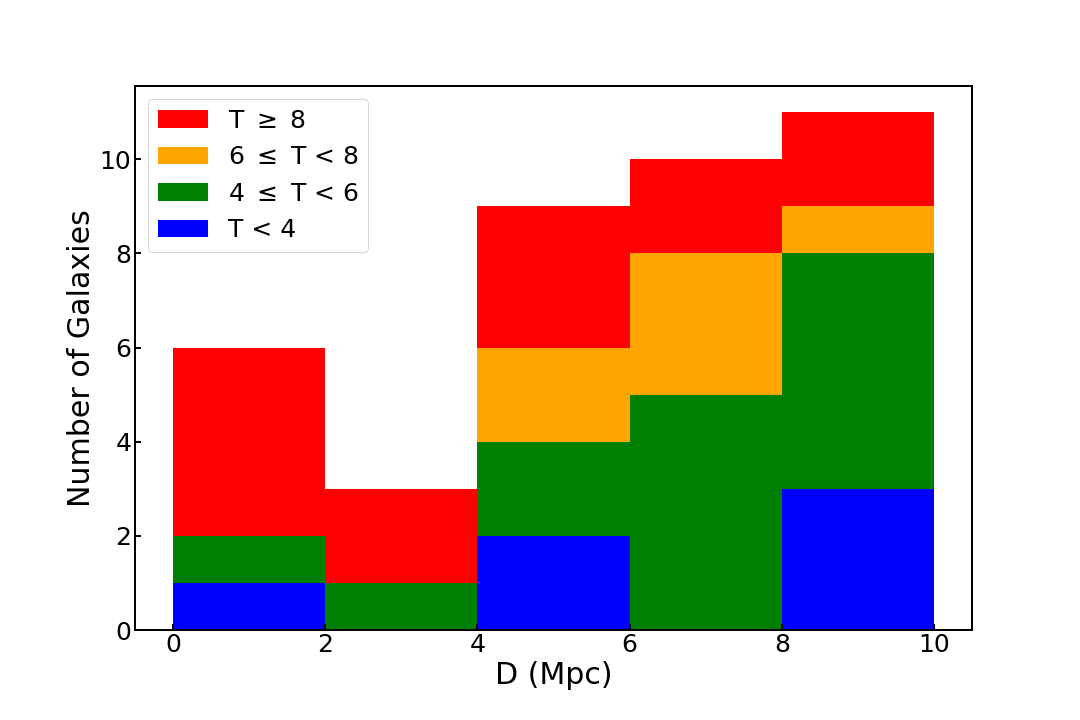}
\vspace{-0.2cm}
\caption{Distribution of the primary targets as a function of distance. The galaxies are color-coded according to their morphological type, as given by the parameter T from the revised Hubble classification scheme.}
\label{figDIST}
\end{figure}

Other selection criteria should also guarantee the uniformity of the sampling of different galactic environments\hc\,\,the oxygen abundance 12+log[O/H], varying from 7.5 in Sextans\,A (\citealt{Kniazev2005}) to $\geq$\,9.0 in objects like M51 and M63 (\citealt{Pilyugin2004a})\scp\,\,the stellar density (proportional to $\mu_{\rm V}$), ranging from $\mu_{\rm V}$ below 30\,mag\,arcsec$^{-2}$ up to 18\,mag\,arcsec$^{-2}$; 
the molecular and neutral gas mass,which is also very different between these galaxies; and galactic structures (size of the bulge, spiral arms, bars, rings, faint external structures, etc.). The objects selected are broadly distributed in mass with the addition of smaller irregular galaxies in order to sample different metallicities and environments with a similar statistical weight. The Hubble type and distance distribution of the sample is shown in Figure\,\ref{figDIST}. The properties that were important for the completion of our sample can be summarized as follows$\hc$

\begingroup{
\leftskip=1.truecm
\parindent=-0.4truecm

1)~global metallicity\hc\ 7.5\,$\lesssim$\,12+log[O/H]\,$\lesssim$\,9.0\scp

2)~magnitude\hc\ $-$21.3\,$\lesssim$\,M$_{\rm absolute}$\,$\lesssim$\,$-$13.5\,mag\scp

3)~surface brightness\hc\ 18\,$\lesssim$\,$\mu_{\rm V}$ $\lesssim$\,30\,mag\,arcsec$^{-2}$\scp~and

4)~galactic environments\hc\ 1/3 isolated objects, 2/3 in groups, with no strongly interacting systems.

}\endgroup
\noindent{\noindent{The group property of the galaxies was extracted from NED\footnote{The NASA/IPAC Extragalactic Database (NED)
is operated by the Jet Propulsion Laboratory, California Institute of Technology,
under contract with the National Aeronautics and Space Administration. }}}

As the data is analyzed, the selection can be adjusted to properly sample the different environments. For example, if we gather enough high metallicity and dense environment regions, we will then concentrate on the small irregular galaxies with a low-metallicity. We will prioritize the completion of the science goals and that depends mostly on the sample of H{\II} regions and less on the completion of one particular target. 

Among SIGNALS'galaxy sample, 7 targets are very close, resulting in a spatial resolution below 3\,pc in the best observing conditions. A mosaic of several fields with small overlaps is used for these objects to cover most of the star-formation activity. Figure~\ref{M33} shows an assemblage of four fields in M33 from the SIGNALS pilot project obtained in 2017. Reaching such a high spatial resolution is essential for some objectives of the project, i.e. to resolve gas ionization structures and the ionizing star content of the nebulae (e.g. faint end of the H{\II} region luminosity function, calibration of the models with resolved structure of ionized gas, etc.). 

The spatial resolution of the sample ranges from 2 to 40\,pc. It should complement the MUSE survey over the northern sky (PHANGS/MUSE). Out of 19 galaxies,  SIGNALS has 6 are in common with PHANGS/MUSE. We also compiled existing data from HST+LEGUS (28 galaxies), GALEX (UV, 39 galaxies), \textit{Spitzer} (IR, 43 galaxies), BIMA and ALMA (CO, 23 galaxies), VLA (HI, 42 galaxies), as well as those obtained with narrow-band filters. These will be used to complete our analysis. The availability of this complementary data for the SIGNALS sample is also indicated in Table\,\,\ref{target_list}.

\section{Science Requirements}

SIGNALS aims to measure the strong emission lines while resolving individual H{\II} regions. We also want to measure the H$\alpha$ brightness of the DIG, study small-scale dynamics of the gas and investigate feedback mechanisms at the scale of H{\II} regions. As described in Section\,\,\ref{str_filter}, the observation strategy was defined from basic constraints related to the detection of faint H{\II} regions and the DIG with respect to their expected surface brightness and the known velocity dispersion observed in the ionized gas. The target selection was based on the spatial resolution requirement and the efficiency at observing a large number of H{\II} regions in different local environments. 

As described in this section, different steps resulting in the proper measurements of the emission line and continuum fluxes, as well as the velocity and its dispersion, were tested during the design of the survey\hc\,\,1)~an additional wavelength calibration for the velocity field is obtained by fitting the OH sky lines of the SN3 filter subtraction\scp\,\,2)~the sky subtraction is performed using sky sampling through each datacube\scp\,\,and 3)~the stellar continuum subtraction is applied pixel by pixel using a model for the stellar population. Finally, once the lines are properly measured, the analysis involves\hc\,\,4)~a dust correction using the Balmer decrement method\scp\,\,5)~the identification of emission regions \scp\,\,and 6)~characterization using photoionization models. 
Requirements for meeting science goals on the study of\hc\,\,7)~the feedback processes\scp\,\,8)~the dynamics\scp\,\,and 9)~the impact of local environments, are also introduced in this section.

\subsection{Wavelength Calibration using Sky Lines}
\label{sec:Star wave}

As mentioned in Section\,\ref{datared}, the wavelength calibration provided by the data reduction pipeline ORBS is performed by measuring the position of the laser line on a He-Ne datacube through the FOV, and applying the measured centroid map to scale the science cube spectral axis (\citealt{Martin2017}). Some velocity residual errors remain from this technique since the science observations are made at different on-sky positions than those of the calibration laser (zenith). As described in \cite{Martin2016} and \cite{Martin2018}, the sky lines observed in the SN3 filter can be used to correct residual errors. A simultaneous fit of all the bright sky lines available in the bandpass is made using the ORCS analysis tool. The resulting velocity map is interpolated over the FOV and a correction map is then applied to any subsequent velocity map extracted from the emission line with ORCS. Note that at R\,$=$\,5000, the sky lines are well resolved and their number reduces the uncertainties on the velocity extracted from the fit. The same accuracy is not required for the SN1 and SN2 filters. 

\subsection{Sky Subtraction}
\label{sec:Star sky}

The sky subtraction is performed 
on each pixel using a high SNR sky spectrum built from each datacube. This sky spectrum is extracted using the median of the available sky spaxels over the FOV. Although a few hundred pixels is enough to produce such a spectrum which corresponds to an area greater then $\sim$10\,arcsec$^{2}$ on the sky, all the available sky spaxels can be used to produce the median sky spectrum. The selection of those sky spaxels is made from the deep image of the datacube (the deep image is the image produced from the stacked datacube along the spectral axis) by defining a maximum threshold on the image intensity. This threshold can be adjusted from one target to another to optimize the selection of the faintest pixels while still using a large number of them. This adjustment is important to minimize the DIG or stellar background contamination to the sky spectrum. We estimate that the vast majority of all SIGNALS fields will have enough sky sampling to perform the high SNR sky spectrum subtraction (i.e. at least 10\,000 sky spaxels corresponding to a SNR $\geq$ 100). For pointings without any pixels enabling the sampling of the sky, we use a model of the sky spectrum obtained by fitting the bright lines with ORCS. Since the emission of the sky can vary in time and since it includes a faint H$\alpha$ component, a model is used only when a high SNR sky spectrum cannot be produced.

\subsection{Stellar Continuum Subtraction}
\label{sec:Star cont}
 
The stellar population contribution to each spaxel is subtracted to obtain the proper emission line fluxes. At the SIGNALS spatial resolution of 2 to 40\,pc, the effect of stellar continuum subtraction is much smaller than at resolutions of 1\,kpc. The emission of the gas is concentrated in clumpy areas while the stellar population in massive galaxies is more uniformly distributed. For example, at a $\mu_{\rm V}$\,$=$\,18\,mag\,arcsec$^{-2}$, the intensity of the continuum is of 2.65$\times$10$^{-17}$ erg\,s$^{-1}$\,cm$^{-2}$\,\AA $^{-1}$\,arcsec$^{-2}$ which corresponds to 2.71$\times$10$^{-18}$ erg\,s$^{-1}$\,cm$^{-2}$\,\AA $^{-1}$\,pixel$^{-1}$. Note that this level corresponds closely to the level of noise in one spaxel for any given SIGNALS datacube. For low-mass galaxies, no continuum subtraction is necessary for our project. Nevertheless, the continuum can dominate the emission in the central part of massive galaxies. To properly measure the emission from faint H{\II} regions or the DIG components in those areas (SB$_{\rm H\alpha DIG}$\,$\simeq$\,5$\times$10$^{-18}$\,erg\,s$^{-1}$\,cm$^{-2}$\,arcsec$^{-2}$), a subtraction of the stellar population background needs to be performed. 

As shown in \cite{Rousseau-Nepton2018}, a stellar continuum can be extracted directly from the datacubes. At a resolution of R\,$=$\,5000, the stellar absorption line can be fitted along with the emission line. This is important for the H$\alpha$ and H$\beta$ absorption lines. For the galaxies observed in SIGNALS, the stellar population continuum subtraction will be performed in three steps\hc\,\,1) a fit of the gas emission will be performed to obtain an estimate of the velocity field, the emission line fluxes, and the continuum level, 2) the aperture for the stellar continuum sample will be defined using the galaxy luminosity profile, 3) the stellar population spectra will be extracted or fitted through the apertures and used to subtract the component from each spaxel (e.g. \citealt{Moumen2019}). Ideally, when the SNR and spatial resolution allow it, the local environment will be used to build the galaxy background spectrum directly from the datacubes. While selecting the pixels to build the stellar continuum spectra, a limit on the combined emission line map flux (from a preliminary fit of the emission with ORCS) is used to exclude pixels with a significant contribution from the DIG and/or surrounding HII regions. If no pixel passes this selection test, or if the noise level of the stellar continuum spectrum is higher than 1/10 of the detection threshold for the HII regions, the aperture is increased. Depending on the galaxy, this method will not always be possible and other ones will be considered.

Also discussed in \cite{Rousseau-Nepton2018}, although some pixels within an emission region may be affected by the young population continuum flux and absorption lines, our method for the subtraction of a local or global background stellar population spectrum takes into account most of the young population emission. Indeed, when the stellar population spectrum is subtracted, it is first rescaled to the continuum level of the emission region, which takes into account the continuum level of the young population. The absorption line profiles in this rescaled stellar population spectrum do not contain the exact contribution from the young population, but this inaccuracy is negligible compared to the final noise level in the emission lines. Tests done with a combination of model spectra for young and old populations with different ages and proportions support this idea.
\subsection{Dust Extinction Correction}
\label{sec:dust}

The dust extinction contribution is included in our model for the spectral analysis. To provide dust corrected fluxes, the extinction law of \citet{Cardelli1989} is used and the color excess ${E(B-V)}$ is evaluated from the known equation\hc\,\,
\begin{equation}
E(B-V)\,=\,\frac{2.5}{1.10} \log \left[ \frac{({\rm F}_{{\rm H}\alpha}/{\rm F}_{{\rm H}\beta})_{\rm obs}}
{({\rm F}_{{\rm H}\alpha}/{\rm F}_{{\rm H}\beta})_{\rm theo}}\right],
\end{equation}
\noindent with $({\rm F}_{{\rm H}\alpha}/{\rm F}_{{\rm H}\beta})_{\rm theo}$ given by the photoionization models, and $E(H\beta-H\alpha)$/$E(B-V)$\,=\,1.10. 

Measured fluxes, both corrected and uncorrected for the dust extinction, will be provided in the catalog of H{\II} regions (see Section\,\ref{dataprod}). 

\subsection{H{\,\sc ii} Region Identification}
\label{sec:Star HII}

\begin{figure}
\centering
\includegraphics[trim={-1cm 0 -1cm 0}, width=3.3in]{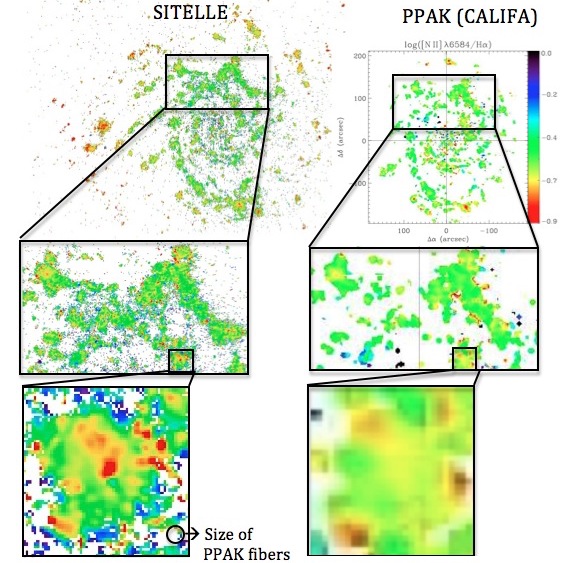}
\caption{Spatial resolution and filling factor impact on the measurements of emission line ratios. The line ratio log([N\II]$\lambda$6583/H$\alpha$) extracted from SITELLE's observations on NGC\,628 (0.8$^{\prime\prime}$ seeing, 0.32$^{\prime\prime}$ per pixel, 100\% filling factor) is compared to PPAK's observations on the same area (0.8$^{\prime\prime}$ seeing, 2.7$^{\prime\prime}$ per fiber, 60\% filling factor). The color scale is the same for both maps (upper-right corner). The PPAK map (on the right) is published in \citet{Rosales-Ortega2011} and was re-used here for comparison purpose only.}
\label{PPAK}
\end{figure}

To separate individual H{\II} regions, a spatial resolution of the order of 10\,pc is required. Figure\,\,9 from \citet{Kollmeier2017} illustrates how the ionized gas structures and individual clumps of star formation can be resolved by increasing the spatial resolution. Below a resolution of 50\,pc, observations reveal structures in the ISM including individual star-forming knots, DIG, and shocks. The filling fraction of the observations are also important for analyzing the gas photoionization conditions. Figure\,\,\ref{PPAK} shows the impact of the FOV and filling-factor on the H{\II} region spectrophotometry by comparing the log([N\II]$\lambda$6583/H$\alpha$) maps obtained with SITELLE and the CALIFA PPAK IFU survey (filling factor $\sim$\,60$\%$, \citealt{Rosales-Ortega2011}). The maps differ significantly. Notably, the flux of faint areas covered by fibers (for fiber-fed systems) becomes negligible resulting in a total flux per resolution element dominated by the brightest zones. Areas not covered by fibers do not contribute to the resulting map. 

\begin{figure*}
\begin{center}
\includegraphics[width=6.9in]{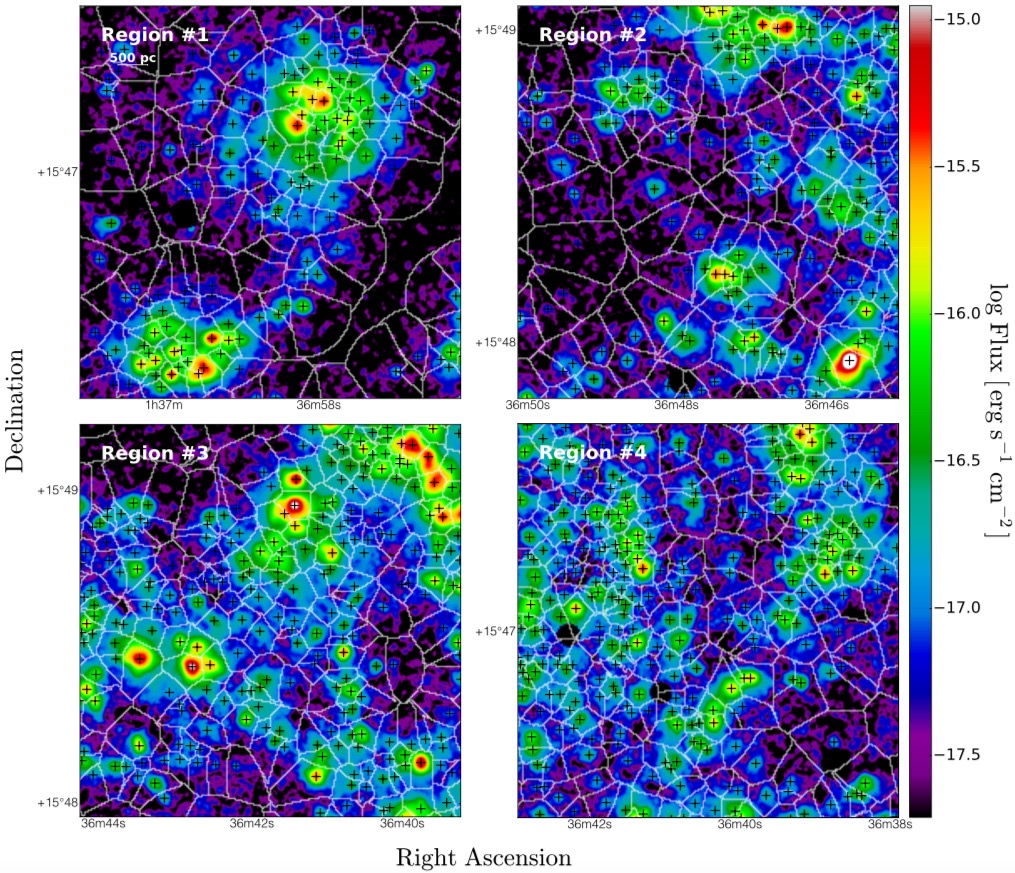}
\caption{Examples of ionizing sources and their zone of influence in NGC\,628 drawn over the H$\alpha$\,+\,H$\beta$\,+\,[O\III]$\lambda\lambda$4959,5007 continuum-subtracted image from \citet{Rousseau-Nepton2018}. The centroid position of each emission peak detected is identified with a cross. The white contours define the zones of influence surrounding the emission peaks.}
\label{HII_detect}
\end{center}
\end{figure*}

Our analysis of the science verification data allowed us to identify 4285\,\,H{\II} region candidates in NGC\,628 using a tailored procedure developed for the SITELLE data at a spatial resolution of 35\,pc. Details are published in \cite{Rousseau-Nepton2018}. There are two advantages of the IFTS that our procedure uses\hc\,\,the full spatial coverage and the spectral information. By combining the flux of three emission lines (H$\alpha$, H$\beta$, and [O\III]$\lambda$5007), we increased the detection threshold and improved the detection of the low metallicity regions with much fainter recombination lines. This code provides flexibility in defining the regions and their DIG background. Our method proceeds through multiple steps\hc\,\,1)the identification of the emission peaks\scp\,\,2) the determination of the zone of influence around each peak\scp\,\,and 3) the definition of the outer limit of a region and its DIG background. Figure~\ref{HII_detect} shows the border of a few H{\II} regions derived from this procedure. For SIGNALS, it will be adapted to different spatial resolutions to make sure that the regions are uniformly defined. SITELLE's dataset for NGC\,628 was used to recover the luminosity function for different samples of H{\II} regions as shown in \cite{Rousseau-Nepton2018}. These luminosity functions, established from 2000\,-\,4000 regions, are well sampled and provide interesting insights on the active star formation processes within that galaxy.

\subsection{H{\,\sc ii} Region Characterization}

Photoionization models are usually employed to understand the physical properties of H{\II} regions, whether or not their ionization structure is resolved, e.g. CLOUDY (\citealt{Ferland1998,Morisset2015}) or MAPPINGS (\citealt{Dopita1995,Ho2014}). The ensemble of emission lines listed in Table~\ref{liner} will allow us to investigate a broad range of physical conditions for the gas and ionizing sources (\citealt{Kewley2002,Perez-Montero2014,Vale-Asari2016}). With those lines (and their ratios), we are planning three different modeling strategies for SIGNALS\hc\,\,Integrated Spectra Modeling, Spatially Resolved Modeling, and Modeling the Populations.

\subsubsection{Emission Line Detection}

The intensities of the strong nebular lines vary with the physical conditions in the gas. Nevertheless, we can make some assumptions about the detectability of the strong lines from previous observations and grids of photoionization models. When a line is not detected, the detection threshold adds a constraint on the line???s flux which can still be used to narrow down the region's physical conditions. Depending on the region size and the spatial resolution, the detection threshold changes and this effect has to be considered as an observational bias in the subsequent analysis. 

\begin{figure}
\begin{center}
\includegraphics[width=3.3in]{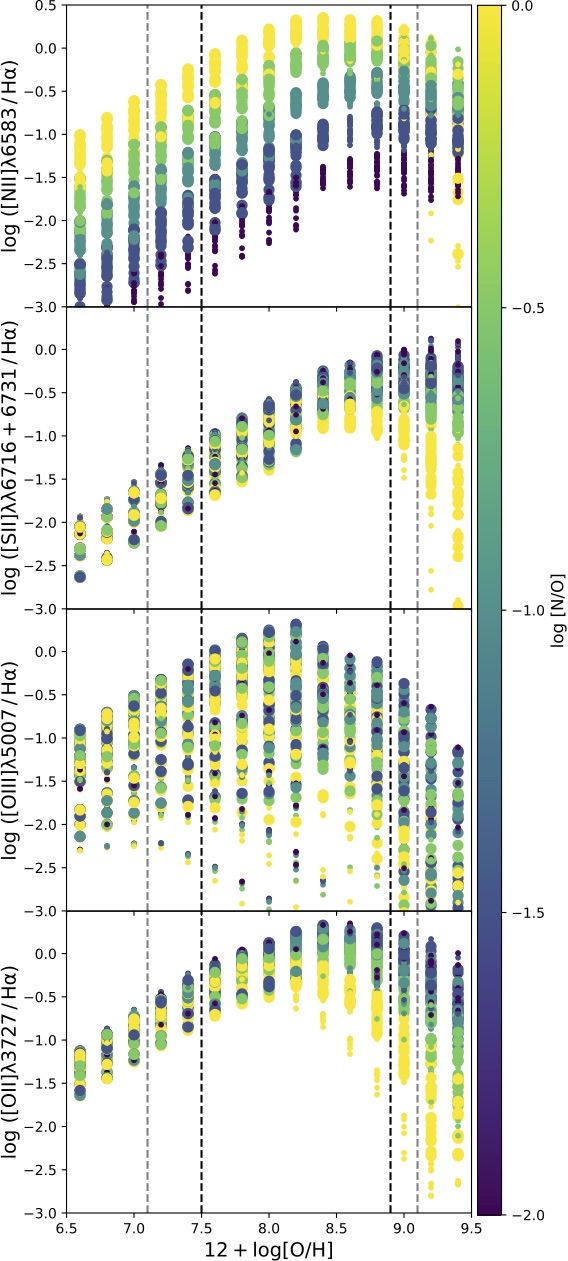}
\caption{Prediction of several strong line ratios ([line]/H$\alpha$) of the CLOUDY photoionization code (extracted from the BOND grid of models on 3MdB). The color scale represents the log[N/O] ratio for a given model. The smaller dots correspond to models representing less expected regions (log[N/O]\,$<$\,$-$1.5 or age\,$>$\,5.1 Myr or log[U]\,$\geq$\,$-$3.55). The grey dashed vertical lines show the expected range of metallicity of the regions sampled and the black dashed vertical line shows the usual range of metallicity for most regions in the local Universe.}
\label{BOND_R}
\end{center}
\end{figure}

For nearby targets, most regions will be spatially resolved and multiple spaxels will be available to increase the detection limit. For the faintest regions, considering a surface brightness of $\sim$\,8\,$\times$\,10$^{-17}$\,erg\,s$^{-1}$\,cm$^{-2}$\,arcsec$^{-2}$ and a radius of 2$^{\prime\prime}$, 
a SNR of 3 would be reached for any line with a flux greater than 12\% of H$\alpha$ (or log[line/H$\alpha$]\,$\geq$\,$-$0.9). For any region with a surface brightness greater than 3.4\,$\times$\,10$^{-16}$\,erg\,s$^{-1}$\,cm$^{-2}$\,arcsec$^{-2}$ and a radius of 2$^{\prime\prime}$, 
a SNR of 3 would be reached for any line with a flux greater than 3\% of H$\alpha$ (or log[line/H$\alpha$]\,$\geq$\,$-$1.5). According to photoionization models (\citealt{Vale-Asari2016}), this corresponds to virtually all the regions for [N\II]$\lambda$6583, [S\II]$\lambda\lambda$6716,6731, and [O\II]$\lambda$3727.  Figure\,\ref{BOND_R} shows the expected ratios of the strong lines over H$\alpha$ for a set of models from the Bayesian Oxygen and Nitrogen abundance Determinations database (BOND, \citealt{Vale-Asari2016}) including models for ages ranging from 1 to 6\,Myr, optical depth of 1.0, log[N/O] ranging from 0 to $-$2.0, 12+log[O/H] ranging from 6.5 and 9.5, and ionization parameters ranging from $-$2.5 to $-$4 (value observed in the CALIFA sample, \citealt{Morisset2016}). We can see that within the range of metallicity expected for the extragalactic H{\II} regions of the sample, all models show strong lines' relative intensities $\geq$\,$-$1.5. For [N\II]$\lambda$6583, we considered only regions with a log[N/O] above $-$1.5, since H{\II} regions are not observed frequently below that threshold (\citealt{Berg2015, Croxall2015, Croxall2016}). It is important to note that for [O\III]$\lambda$5007, some cases will not detected, i.e. regions older than 5\,Myr or with a low ionization parameters ($\leq$\,$-$3.5). Note that in both cases the regions are fainter compared to others since the ionizing photons come from late B star(s). Nevertheless, not detecting the [O\III]$\lambda$5007 line is a good constraint for the models. From the whole sample of spatially resolved H{\II} regions (2-15\,pc), a good fraction of them will have strong emission lines detected along the region profile, useful for our detailed 2D analysis (see Section\,\ref{srm}).

\begin{figure}
\begin{center}
\includegraphics[width=3.3in]{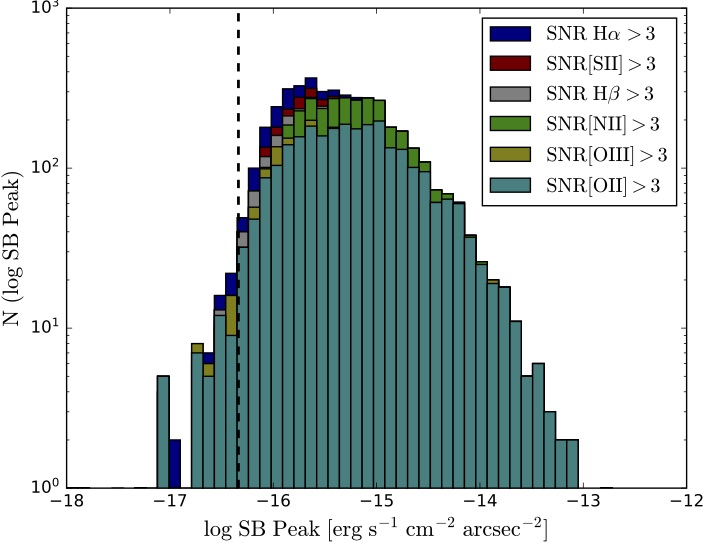}
\caption{Surface brightness (SB) function of the H{\II} regions in the commissioning data of NGC\,628. The black dashed line shows the detection threshold of one element of resolution in H$\alpha$. The different colors represent selected subsets of regions for which there is a detection of different strong lines (see the legend). The few regions below the black dashed line are faint H$\alpha$ regions detected from the combination of H$\alpha$, H$\beta$ and [O\III]$\lambda$5007 used for the identification.}
\label{SB_plot}
\end{center}
\end{figure}

For further-away targets, we can approximately estimate the number of regions for which we will detect strong emission lines by comparing model predictions with observations. Two parameters mainly influence the detection of emission lines for unresolved regions$\hc$\, the size distribution of the regions with respect to the resolution of the data, and the SB distribution of regions with respect to their size. Some small and faint regions cannot be detected in targets near the distance limit of the SIGNALS sample. Their flux is diluted in the resolution element. In the commissioning data of NGC\,628, located at 9\,Mpc, our measured detection threshold for H$\alpha$ was of 4.6\,$\times$\,10$^{-17}$\,erg\,s$^{-1}$\,cm$^{-2}$\,arcsec$^{-2}$ (slightly lower than for SIGNALS). Figure\,\ref{SB_plot} shows the measured peaks' SB function (histogram of the SB of the central pixel) of the regions detected with the identification procedure (\citealt{Rousseau-Nepton2018}) for different subsets of regions on NGC\,628 for which there is a detection of a different strong line. From the regions detected with H$\alpha$, 84\% were also detected in [N\II]$\lambda$6583, 91\% in [S\II]$\lambda\lambda$6716+6731, 63\% in [O\III]$\lambda$5007, 84\% in H$\beta$, 63\% in [O\III]$\lambda$5007, and 63\% in [O\II]$\lambda$3727. Part of the lower detection of the [O\II]$\lambda$3727 line is due to dust extinction, a total exposure time of 2.13 hours (lower than for SIGNALS), and the loss of reflectivity of the telescope primary mirror (only a few months before realuminizing). For [O\III]$\lambda$5007, about 5\% of the regions were at a metallicity too high to be detected and, as mentioned before in this section, the other non-detected regions could be older than 5\,Myr and/or they have a low ionization parameter. From the distance of NGC\,628 to the closest target of the sample, the fraction of detected lines increases as more pixels are available to extract the spectra. NGC\,628 is therefore a worst-case scenario for the detection of the lines. 

\subsubsection{Integrated Spectra Modeling}

\begin{figure}
\centering
\includegraphics[trim={0.4cm 0.3cm 0.4cm 0.3cm}, clip, width=3.25in]{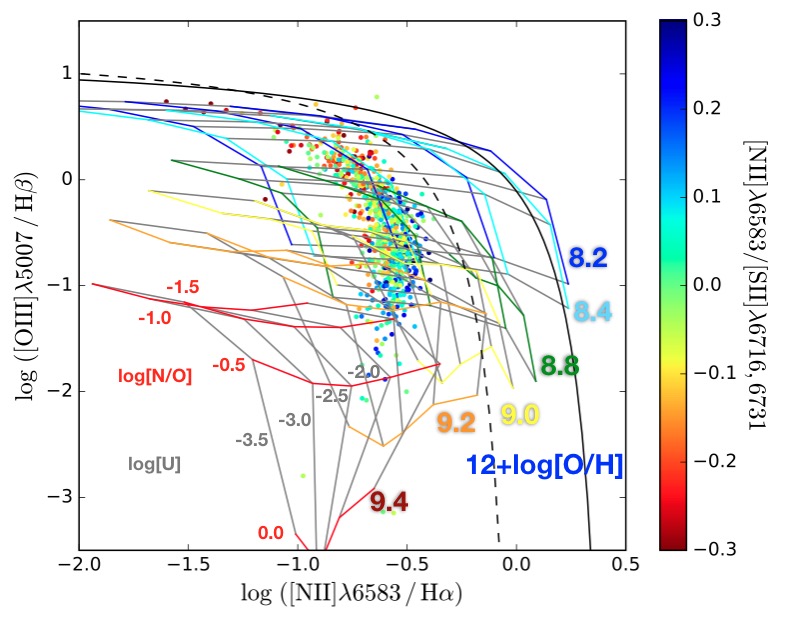}
\caption{BPT diagram (\citealt{BPT1981}) of emission line regions identified in NGC\,628 (from \citealt{Rousseau-Nepton2018}) along with a grid of models from BOND (\citealt{Vale-Asari2016}) using the SED of a 3\,Myr old stellar cluster with a Chabrier IMF (\citealt{Chabrier2003}). The colors of the grids indicate different metallicities (12+log[O/H]). 
Horizontal lines have different values of log[N/O], as given in red for 12+log[O/H]\,=\,9.4, and
vertical lines have different values of log[U], as given in gray for the same metallicity. The color-scale of the scattered observational points indicates the log([N\II]/[S\II]$\lambda\lambda$6716+6731) line ratio which is known to be a good indicator of log[N/O] (\citealt{Perez-Montero2014}). The limit between the H{\II} regions and the transition regime (\citealt{Kauffmann2003}) is shown by the dashed curve, and the extent of the starburst domain (\citealt{Kewley2001}), is displayed by the continuous curve.}
\label{models}
\end{figure}

Grids of models covering different physical properties from various photoionization codes (BOND, \citealt{Vale-Asari2016}\scp\,\,3MdB, \citealt{Morisset2015}\scp\,\,and MAPPINGS, \citealt{Dopita1995}) will be compared to the SIGNALS' H{\II} regions integrated spectra. These grids are designed to study the effect of variations in the gas conditions on the integrated spectrum (region by region). Figure~\ref{models}
shows an example of the comparison of emission line ratios observed in NGC\,628 (\citealt{Rousseau-Nepton2018}) with BOND. 

Different models can account for variations in the spectral energy distribution (SED) of the source, the metallicity [O/H], and the relative abundances of elements (e.g. [N/O]). The accuracy of the comparison with models can be improved by using a set of regions for which the ionization sources have been observationally constrained, e.g. Legacy ExtraGalactic UV Survey (\href{https://legus.stsci.edu}{LEGUS}, \citealt{Calzetti2015, Grasha2015}). LEGUS provides access to both individual stars (\citealt{Sabbi2018}) and stellar cluster catalogues (\citealt{Adamo2017}). It also contains a fairly complete census of the massive stars (down to 15\,M$_\odot$) and very young clusters (down to 10$^{3}$\,M$_\odot$). Thus, the combined data from SIGNALS and LEGUS will provide a better view of the H{\II} region physics for a large fraction of our sample.

\subsubsection{Spatially Resolved Modeling}
\label{srm}

When a H{\II} region is resolved, spatial variations of the ratios as a function of the distance to the source of ionization are expected. This can be due to the varying photoionization conditions in the gas as the intensity and slope of the ionizing spectrum, temperature, and density change. Once integrated, the variations can be less apparent (e.g. [NII]/[OII] ratio, see Figures on p.37 of Rousseau-Nepton at al. 2018). 

\begin{figure}
\begin{center}
\includegraphics[trim={0.cm 0.cm 0.cm 0.cm}, clip, width=3.3in]{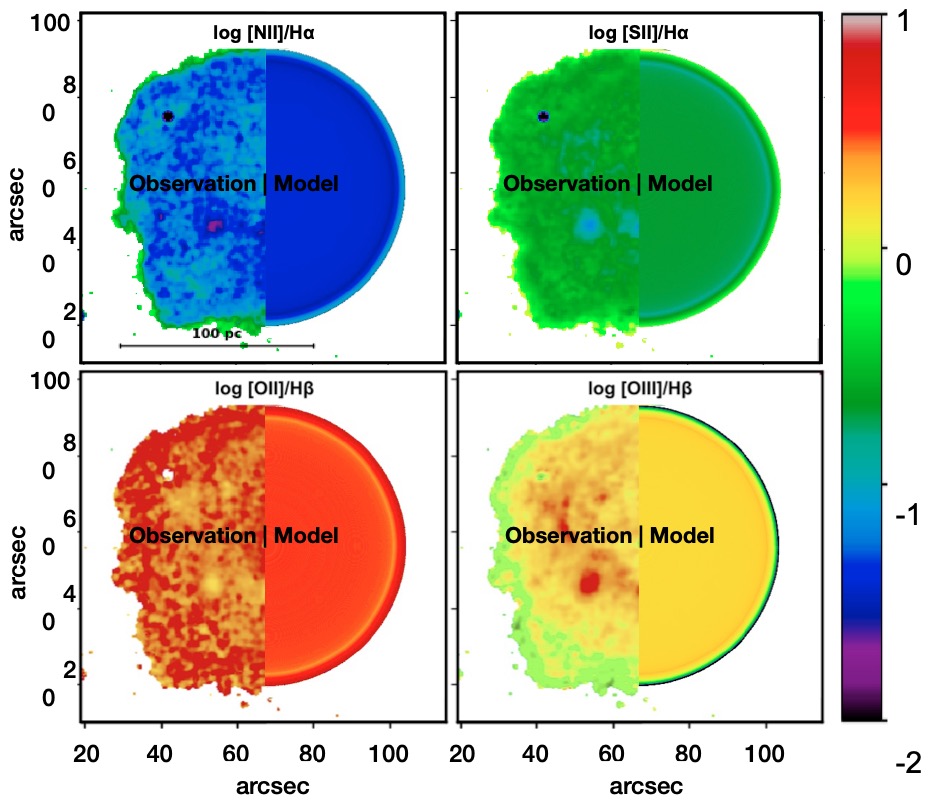}
\vspace{-0.5cm}
\caption{Comparison of a giant H{\II} region observed with SITELLE in NGC\,6822 with a 3D photoionization model projected on 2D images. The region is split in two for each panel\hc\,\,the left side represents the observations while the right side is the model (a slice perpendicular to the line of sight that goes from the center of the region to the outskirts). A pure 40\,000\,K Planck function was used as ionizing spectrum and abundances at 20\% of the solar value, excepted for S. The model was also for a thin shell of gas with log[U]\,=\,$-$2.6.}\label{model_chris}\end{center}
\vspace{-0.5cm}
\end{figure}

The 2D sampling of the line emission enables the use of multi-dimensional photoionization models to constrain gas properties (\citealt{Wood2005,Wood2013,Weber2015,Rahner2019}). For each source identified, we establish a mean profile of the line ratios as a function of the distance to the ionization source (OB stars in the case of H{\II} regions). These profiles along with the integrated flux of each line are needed to conduct a fair comparison with the models. Figure~\ref{model_chris} shows a comparison of four line ratios with a 2D projection of the pseudo-3D model extracted from pyCLOUDY (\citealt{Morisset2013}). 

\subsubsection{Modeling Populations}

\begin{figure}
\centering
\includegraphics[trim={7.2cm 2.5cm 1.6cm 2.8cm}, clip, width=0.98\linewidth]{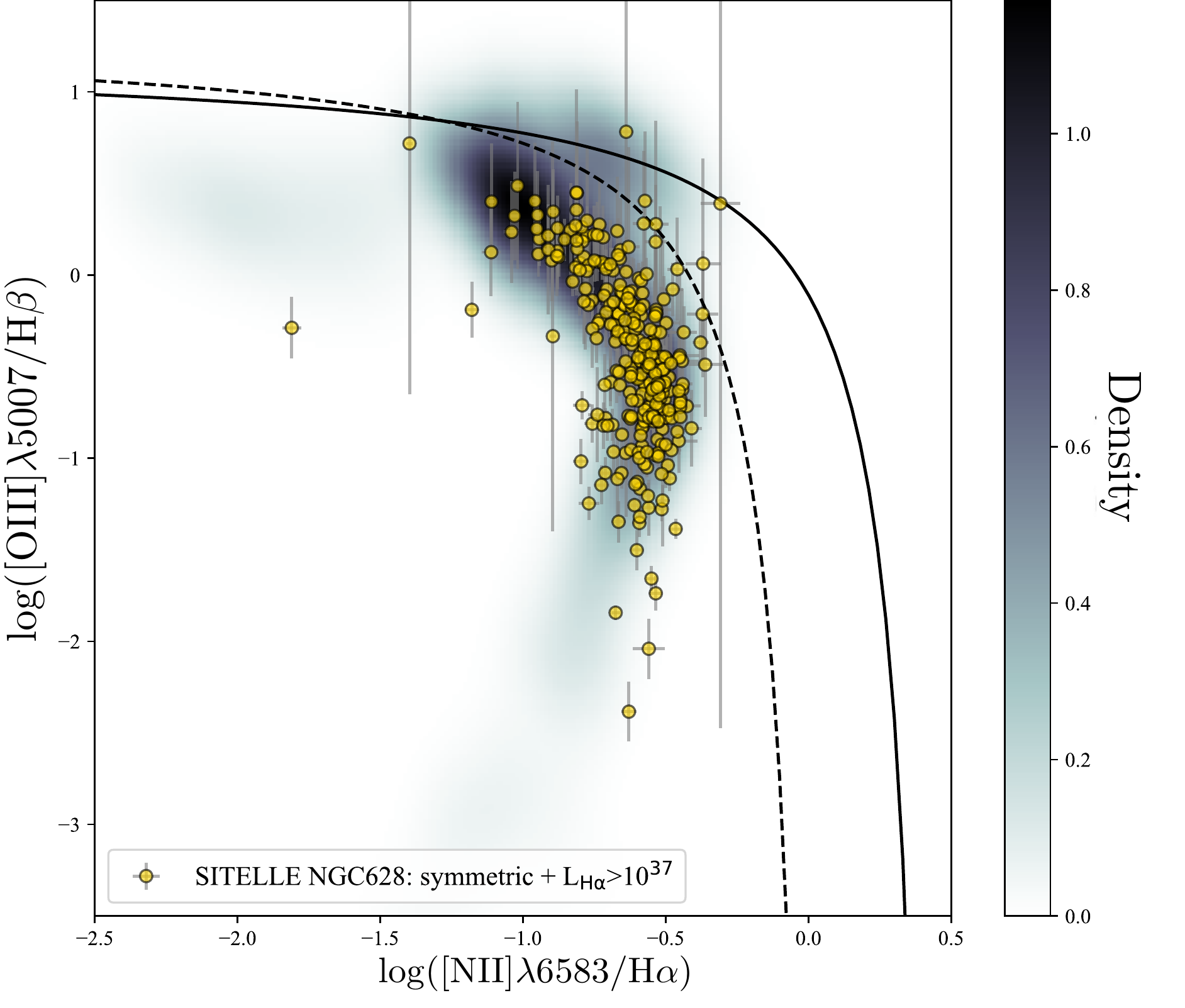}
\vspace{-0.2cm}
\caption{BPT diagram (\citealt{BPT1981}) of emission line regions identified in NGC\,628 from \citet{Rousseau-Nepton2018}, using circular regions only, along with simulations from CLOUDY using post-processing of evolutionary tracks from the WARPFIELD code (\citealt{Rahner2019, Pellegrini2019}). It combines a series of WARPFIELD runs with different cloud masses, densities, and ages. The color-coding indicates the relative number of model outputs used to produce
a line ratio. See the caption of Figure~\ref{models} for the curve definitions.}
\label{model_germany2}
\end{figure}

In addition to the region-by-region analysis, a statistical approach is used to study populations of H{\II} regions. Figure~\ref{model_germany2} shows simulations of line ratios when combining photoionization models with evolutionary tracks of a population of star-forming clusters at solar metallicity. The emission measure predictor code \href{https://bitbucket.org/drahner/warpfield}{WARPFIELD} (Winds And Radiation Pressure\hc\,\,Feedback Induced Expansion, colLapse and Dissolution, \citealt{Rahner2017,Rahner2019, Pellegrini2019}) was used to produce Figure~\ref{model_germany2}. WARPFIELD considers a semi-analytic model to describe the impact of mechanical and radiative feedback from a young massive cluster on its parental cloud. Its approach simultaneously and self-consistently calculates the structure and the expansion of shells driven by feedback from stellar winds, supernovae, and radiation pressure, while accounting for the deceleration of the shell due to gravity. The model has been used to investigate the conditions under which the different sources of feedback dominate and the amount of radiation that escapes through the shell. It was also used to derive the minimum star formation efficiency for a large parameter space of clouds and clusters (\citealt{Rahner2017}). The same approach was used to explain the two stellar populations in NGC\,2070 in the LMC, which hosts the younger cluster R136 in its center (\citealt{Rahner2018}). Because this method is computationally very efficient, it will allow us to consider a large range of parameters. We are currently compiling a comprehensive database where we vary the star formation efficiency, the cloud mass and density profile, and the metallicity of both the gas and the stars. Each model is currently being post-processed using CLOUDY to make detailed predictions for the time evolution of the line and continuum emission associated with the system. Using this code, we can build synthetic BPT-like diagrams for the one-to-one comparison with observational data, as shown in Figure~\ref{model_germany2}. 

\subsection{Feedback Processes}
\label{feedback}

Massive OB~stars in H{\II} regions are not only responsible for ionizing the surrounding gas but also for exerting direct radiation pressure onto this gas. They also return enriched material and mechanical energy through their stellar winds and supernova explosions to the ISM. While the efficiency of stellar winds involved in shaping an H{\II} region bubble is still uncertain, it is possible that radiation pressure may play a dominant role at a younger age, while thermal pressure of the warm ionized gas may become more important at a later age (\citealt{Lopez2014,Pellegrini2019}). Also, the DIG can be observed near H{\II} region boundaries. Recent works on the DIG have helped to quantify its emission in nearby galaxies (\citealt{Lacerda2018,Poetrodjojo2019,Moumen2019}) and photoionization models have been developed to trace back the spectral energy distribution of the source of ionization. Several sources of ionization might be responsible for the DIG: ionizing photons escaping H{\II} regions that travel long distances (\citealt{Ferguson1996,Haffner2009, Zurita2000, Oey2007, Barnes2014, Howard2016, Howard2017b}), a weak AGN (\citealt{Ho2008,Davies2014}), a generation of post-AGB stars (\citealt{binette1994, Flores-Fajardo2011}), and fast shocks in the ISM (\citealt{Allen2008,Hoffmann2012}). Some studies have suggested that only OB stars could be responsible for most of the DIG emission (\citealt{Domgorgen1994}). Others have found that a mix of multiple sources could explain the observed line ratios in the DIG (\citealt{Weber2019,Poetrodjojo2019}). 

Detailed numerical simulations predict that the fraction of escaping photons varies considerably over a few million year timescale. Semi-analytic feedback models for the cloud dissolution (\citealt{Rahner2017}) have demonstrated that the SFE is linked to the ionizing escaping photons (i.e. a high SFE corresponds to a high fraction of escaping photons). Also, depending on the evolution phase of the H{\II} region, the escaping photon fraction can change (\citealt{Pellegrini2019}). SIGNALS will measure the DIG surrounding each H{\II} region as shown in \cite{Rousseau-Nepton2018}. By measuring the relative contribution of the H{\II} regions to their surrounding DIG, we will get precise constraints on the fraction of radiation and hot gas escaping the regions that merges with the low density ISM. 

Our requirements for the study of feedback processes rely on\hc\,\,1)~the spatial resolution to resolve regions, filaments, supernova remnants, etc.\scp\,\,2)~the depth of the observations to detect faint regions and DIG emission\scp\,\,and 3)~the line ratios to track the ionizing front or its absence around H{\II} regions along with the variation of chemical abundances in the gas. These requirements have already been met by the need to identify and characterize H{\II} regions. 

\subsection{Kinematics and Dynamics}
\label{sec:small-scale-dynamics}

\begin{figure}
\begin{center}
\includegraphics[width=3.3in]{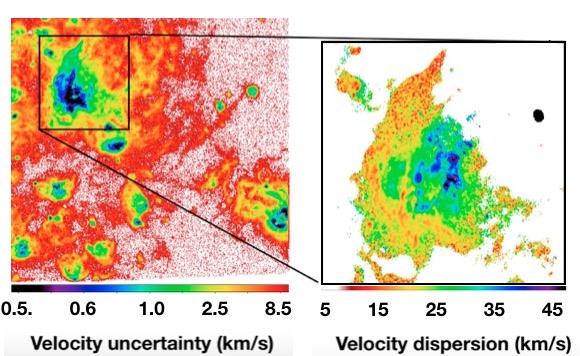}
\vspace{-0.4cm}
\caption{M33 H$\alpha$ velocity uncertainties and dispersion for a small portion of the SITELLE FOV with $R\,=\,3\,000$.}
\vspace{-0.5cm}
\label{velodis}
\end{center}
\end{figure}

The spectral resolution provided by SITELLE SN3 cubes for five lines simultaneously (R\,=\,5\,000 for the H$\alpha$, [N\II]$\lambda\lambda$6548,6583, and [S\II]$\lambda\lambda$6717,6731 emission lines) will allow us to determine the line centroid (ORCS\scp\,\,see Section\,\ref{datared}) with a precision of 0.1 to $\sim$10\,km\,s$^{-1}$ across the FOV. This will enable the detection of line broadening larger than 10\,km\,s$^{-1}$ in most H{\II} regions (see Figures~\ref{vel_snr} and \ref{velodis}). It will also separate multiple components along the line of sight (\citealt{Martin2016}). Combined with the line ratio analysis discussed in the previous subsections, several aspects of the ionized gas dynamics will be probed\hc\,\,1)~shell expansion velocity in H{\II} regions, from which their mass can be determined\scp\,\,2)~random and rotational motions as a function of location in the disk (e.g. spiral arms versus interarm regions and DIG)\scp\,\,3)~azimuthal anisotropy of random motions in the disk plane\scp\,\,4)~kinematics and abundance patterns linked to the mixing mechanisms and abundance variations \citep{sanchezmenguiano2016}\scp\,\,5)~kinematics of the ionized gas compared\scp\,\,and 6)~mass distribution models \citep{kam2017}. In a few objects, the kinematics of the stellar halo should also be probed through the [O\III]$\lambda$5007 emission line in PNe, allowing alternative derivations of rotation curves and mass profiles.

\begin{figure*}
\begin{center}
\includegraphics[trim={0.1cm 0.5cm 0.1cm 1.5cm}, clip, width=6.9in]{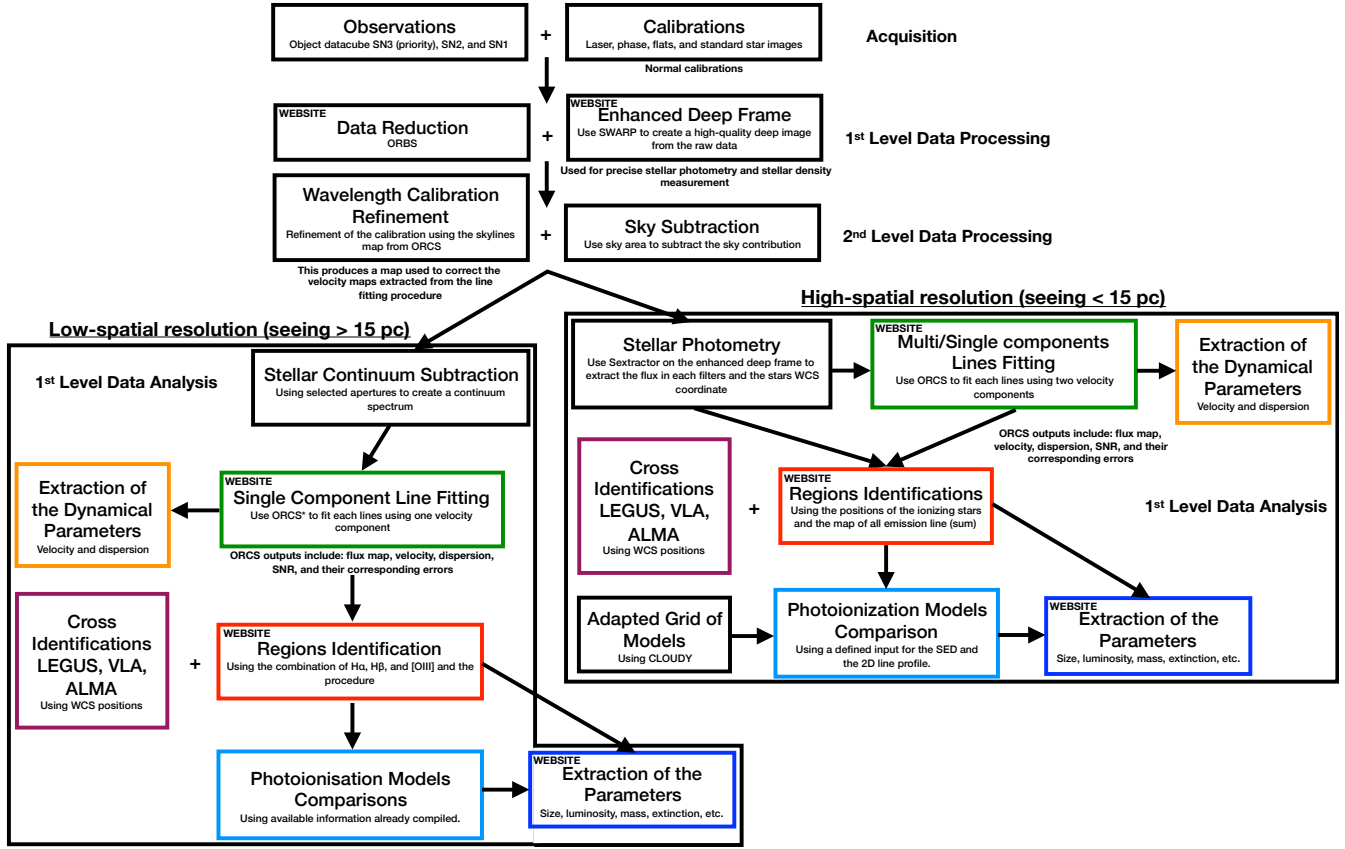} 
\vspace{-0.2cm}
\caption{Organizational chart for the data processing. All items with the label WEBSITE will be accessible to everyone via our \href{http://www.signal-survey.org}{Website} (http://www.signal-survey.org) once the collaboration team has analyzed the data and published the different catalogs. Note that all observed datacubes are available at the Canadian Astronomical Database Center  (\href{http://www.cadc-ccda.hia-iha.nrc-cnrc.gc.ca/en/search/?collection=CFHT&Observation.instrument.name=SITELLE&noexec=true}{CADC}).}
\vspace{-0.4cm}
\label{process}
\end{center}
\end{figure*}

\subsection{Impact of Local Environment}\label{local-env}

To investigate possible relations between star-forming regions and their environment, the following parameters will be studied\hc\,\,the local stellar density, local DIG background, local neutral gas density, local SFR and SFE, distances to the nearest regions, and location with regards to prominent galactic structures (bar, spiral arm, ring, AGN, etc). The size, geometry, and total H$\alpha$ luminosity will also be considered. A large variety of galaxies with different stellar populations and stellar densities can be found within the SIGNALS sample. Within each individual target this distribution also changes dramatically from one area to another. While the diversity of targets adds complexity to the analysis, it makes the analysis robust because reliable subsamples can be used for specific purposes in addressing the main goals of the project. 

\section{Data Products}
\label{dataprod}

Figure~\ref{process} summarizes the diverse processing steps presented in this paper, from the observational phase to the extraction of the physical parameters. The process differs between nearby (i.e. spatial resolution $\lesssim$\,15\,pc) and more distant targets. The ORCS software can simultaneously or individually fit lines in each datacube while returning maps for each line including the amplitude, FWHM, continuum height, flux, velocity, velocity dispersion, and their corresponding uncertainties. The H{\II} regions identified with our procedure will be included in a catalog that will contain the raw emission line measurements from the datacubes obtained after data processing and the stellar continuum subtraction. The median profile of each line will also be fitted as in \cite{Rousseau-Nepton2018}; the fitting parameters will be provided in the catalog. Extinction corrected values will also be provided as explained in Section\,\ref{sec:dust}. The absolute position (right ascension and declination) of the center of each region will be provided as well as the name of the host galaxies. Additional data products extracted from complementary data will be added to the catalog when available (e.g. the mean $\mu_{\rm V}$ around the regions, and the total HI and CO emission evaluated over the region radius). \\
\indent Data products will be distributed to the collaborators and other communities from the SIGNALS \href{http://www.signal-survey.org}{Website}. The HDF5 files containing the reduced datacubes are available without any proprietary period from the program 18BP41, 19AP41, ...P41, etc, via the Canadian Astronomy Data Centre \href{http://www.cadc-ccda.hia-iha.nrc-cnrc.gc.ca/en/}{Website}\footnote{http://www.cadc-ccda.hia-iha.nrc-cnrc.gc.ca/en/}. Data release of additional products following the analysis will be made available later from the SIGNALS \href{http://www.signal-survey.org}{Website}.

\section{Legacy Value}
\label{sec:legacy}

SIGNALS fills the gap between other galaxy surveys focusing on the local Universe, from the upcoming SDSS-V LVM survey to current surveys such as MAD, PHANGS, CALIFA, SAMI, MANGA, etc. It will provide new, key insights into star formation and feedback mechanisms driving the evolution of the ISM in star-forming galaxies. By using multiple nebular lines, physical properties like abundances, ionization structures, and dynamics will be obtained as well as information on the impact of local environments on the star formation process. 

The rich SIGNALS dataset will be valuable for investigating other complementary astrophysical topics. Some possibilities include\hc\,\,1)~planetary nebula abundance distributions and luminosity functions (\citealt{Kreckel2017,Martin2018})\scp\,\,2)~supernova remnant ionization conditions, occurrence, and feedback contribution (Moumen et al. 2019)\scp\,\,3)~detection of background emission line objects (e.g. [O\II] and Ly-$\alpha$ emitters)\scp\,\,and 4)~large-scale velocity mapping of galaxies. The SIGNALS collaboration team will make this dataset available to the scientific community, provide catalogs and diagnostic tools.

\section*{Acknowledgements}

This research is based on observations obtained at the Canada-France-Hawaii Telescope (CFHT) which is operated from the summit of Mauna Kea by the National Research Council of Canada, the Institut National des Sciences de l???Univers of the Centre National de la Recherche Scientifique of France, and the University of Hawaii. The authors wish to recognize and acknowledge the very significant cultural role that the summit of Maunakea has always had within the indigenous Hawaiian community. We are most grateful to have the opportunity to conduct observations from this mountain.
The observations were obtained with SITELLE, a joint project between Universit\'e Laval, ABB-Bomem, Universit\'e de Montr\'eal, and the CFHT, with funding support from the Canada Foundation for Innovation (CFI), the National Sciences and Engineering Research Council of Canada (NSERC), Fonds de Recherche du Qu\'ebec - Nature et Technologies (FRQNT), and CFHT.
The collaboration is grateful to\hc\,\,the Fonds de recherche du Qu\'ebec - Nature et Technologies (FRQNT), CFHT, the Canada Research Chair program, the Natural Sciences and Engineering Research Council of Canada (NSERC), the Swedish Research Council (Vetenskapsr??det), the Swedish National Space Board (SNSB), the Royal Society and the Newton Fund via the award of a Royal Society--Newton Advanced Fellowship (grant NAF\textbackslash{}R1\textbackslash{}180403), FAPESC, CNPq, FAPESP (project 2014/11156-4), FAPESB project 7916/2015, project CONACyT-CB2015-254132. 
\pagebreak

\bibliographystyle{mnras}
\bibliography{SIGNALS_LP_MNRAS} 
\clearpage

\noindent \newline
\textbf{AFFILIATIONS} \smallskip \newline
$^{1}$ Canada-France-Hawaii Telescope, Kamuela, HI, United States \newline
$^{2}$ Department of Physics and Astronomy, University of Hawaii at Hilo, Hilo, HI, United States \newline
$^{3}$ D\'epartement de physique, de g\'enie physique et d'optique,
Universit\'e Laval, Qu\'ebec, QC, Canada\newline
Centre de Recherche en Astrophysique du Qu\'ebec \newline
$^{4}$ Aix Marseille Univ, CNRS, CNES, LAM, Marseille, France \newline
$^{5}$ Department of Astronomy, The Oskar Klein Centre, Stockholm University, Stockholm, Sweden \newline
$^{6}$ Instituto de Astronomia, Universidad Nacional Autonoma de Mexico, Apdo. postal 70???264, Ciudad Universitaria, Mexico CDMX 04510, Mexico \newline
$^{7}$ Department of Physics and Astronomy, University of Western Ontario, London, ON, Canada \newline
$^{8}$ Institute for Astronomy, University of Hawaii, Honolulu, HI, United States \newline
$^{9}$ 2Sub-department of Astrophysics, University of Oxford, Denys Wilkinson Building, Keble Road, Oxford, United Kingdom \newline
$^{10}$ Centro de Astronom\'ia, Universidad de Antofagasta, Avda. U. de Antofagasta 02800, Antofagasta, Chile \newline
$^{11}$ Departamento de F\'isica???CFM, Universidade Federal de Santa Catarina, Florian\'opolis, Santa Catarina, Brazil \newline
$^{12}$  LERMA, Observatoire de Paris, PSL Research Univ., CNRS, Universit\'e de Sorbonne, UPMC, Paris, France \newline
$^{13}$ Coll??ge de France, 11 Pl Marcelin Berthelot, 75005 Paris \newline
$^{14}$ Institute of Astronomy, the University of Tokyo, Tokyo, Japan \newline
$^{15}$ Universit\'{a}t Heidelberg, Zentrum f\"{u}r Astronomie, Heidelberg, Germany  \newline
$^{16}$ Observatoire de Gen\`eve, Universit\'e de Gen\`eve, Sauverny, Switzerland \newline
$^{17}$ Instituto de Astronomia, Universidad Nacional Autonoma de Mexico, Ensenada, B.C., Mexico \newline
$^{18}$ Max Planck Institute for Astronomy, Heidelberg, Germany \newline
$^{19}$ Research School of Astronomy and Astrophysics, Australian National University, Canberra, Australia \newline
$^{20}$ ARC Centre of Excellence for All Sky Astrophysics in 3 Dimensions, Australia \newline
$^{21}$ University of Victoria, Victoria, BC, Canada \newline
$^{22}$ NRC Herzberg Institute of Astrophysics, Victoria, BC, Canada \newline
$^{23}$ D\'epartement de Physique, Universit\'e de Montr\'eal, Montr\'eal, QC, Canada \newline
$^{24}$ Instituto de Astrof\'isica de Andaluc\'ia - CSIC, Granada, Spain \newline
$^{25}$ Estaci\'on Experimental de Zonas \'Aridas, Almer\'ia, Spain \newline
$^{26}$ Observatoire d???Astrophysique de l???Universit?? de Ouagadougou, Ouagadougou, Burkina Faso \newline
$^{27}$ Department of Astronomy, University of Cape Town, Cap Town, South Africa \newline
$^{28}$ Steward Observatory, University of Arizona, Tucson, AZ, United States \newline
$^{29}$ Institute for Astronomy, Astrophysics, Space Applications \& Remote Sensing, National Observatory of Athens, P. Penteli, 15236, Athens, Greece
$^{30}$ Department of Astronomy \& Astrophysics, University of Toronto, Toronto, ON, Canada \newline
$^{31}$ Department of Physics, University of Warwick, Coventry, United Kingdom \newline
$^{32}$ Department of Astronomy, University of California Berkeley, Berkeley, CA 94720, USA \newline
$^{33}$ Department of Physics and Astronomy, Texas Tech University, PO Box 41051, Lubbock, TX 79409, USA \newline
$^{34}$ Departamento de Investigaci\'oon B\'asica, CIEMAT, Madrid, Spain \newline
$^{35}$ Department of Astronomy, School of Science, The University of Tokyo, Tokyo, Japan \newline
$^{36}$ Institute of Astronomy, National Tsing Hua University, Hsinchu, Taiwan \newline
$^{37}$ Academia Sinica, Institute of Astronomy \& Astrophysics, Taipei, Taiwan \newline
$^{38}$ Indian Institute of Astrophysics, Bangalore, India \newline
$^{39}$ Department of Physics and Astronomy, State University of New York at Geneseo, Geneseo, NY, United State \newline
$^{40}$ Departamento de F\'isica Te\'orica y del Cosmos, Universidad de Granada, Facultad de Ciencias (Edificio Mecenas), Granada, Spain \newline
$^{41}$ Instituto Universitario Carlos I de F\'isica Te\'orica y Computacional, Universidad de Granada, Granada, Spain \newline
$^{42}$ Institute for Astronomy, University of Hawaii, HI, United States \newline
$^{43}$ Laborat\'orio de Astrof\'isica Te\'orica e Observacional, Universidade Estadual de Santa Cruz, Ilh\'eus, Bahia, Brazil \newline
$^{44}$ Instituto de Astrof\'isica de Canarias, Tenerife, Spain \newline
$^{45}$ Departamento de Astrof\'isica, Universidad de La Laguna, Tenerife, Spain \newline
$^{46}$ Department of Physics and Space Science, Royal Military College of Canada, Kingston, Ontario, Canada \newline
$^{47}$ LUTH, CNRS, Observatoire de Paris, PSL University, Meudon, France \newline
$^{48}$ School of Physics and Astronomy, University of St Andrews, North Haugh, St Andrews KY16 9SS, UK  \newline
$^{49}$ Royal Society-Newton Advanced Fellowship \linebreak

\appendix

\begin{landscape}

\section{Sample Table}
\begin{table}
\caption{List of targets for SIGNALS}\label{target_list}
\small
\begin{tabular}{|c|c|c|c|c|c|c|c|c|c|c|c|c|c|c|c|c|c|c|}
\hline
\small \makecell{%
  ID \\
  (1)
} & \small \makecell{%
  RA \\
  (2)
} & \small \makecell{%
  DEC \\
  (3)
} & \small \makecell{%
  Morphology \\
  (4)
} & \small \makecell{%
  D \\
  (5)
} & \small \makecell{%
  m \\
  (6)
} & \small \makecell{%
  M \\
  (7)
} & \small \makecell{%
  a \\
  (8)
} & \small \makecell{%
  b \\
  (9)
} & \small \makecell{%
  \textit{i} \\
  (10)
} & \small \makecell{%
  Z \\
  (11)
} & \small \makecell{%
  Ref$\_$Z \\
  (12)
} & \scriptsize \makecell{%
  PHANGS \\
  \small (13)
} & \scriptsize \makecell{%
  LEGUS \\
  \small (14)
} & \scriptsize \makecell{%
  HST \\
  \small (15)
} & \scriptsize \makecell{%
  GALEX \\
  \small (16)
} & \small \makecell{%
  \textit{Spitzer} \\
  \small (17)
} & \scriptsize \makecell{%
  CO \\
  \small (18)
} & \scriptsize \makecell{%
  VLA \\
  \small (19)
} \\
\hline 
\small WLM & \small 00h01m58,16s & \small $-$15d27m39,3s & \small IB(s)m & \small 1.03 & \small 11.03 & \small $-$14.93 & \small 11.5 & \small 4 & \small 70 & \small 7.74 & \small 2 & \small  & \small  & \small  & \small \small x & \small x & \small  & \small x \\ 
\hline  
\small M31 & \small 00h42m44,35s & \small +41d16m08,6s & \small SA(s)b & \small 0.78 & \small 4.36 & \small $-$21.20 & \small 190 & \small 60 & \small 71 & \small 8.72 & \small 1 & \small  & \small  & \small x & \small x & \small  & \small x & \small x \\ 
\hline  
\small NGC\,247 & \small 00h47m08,55s & \small $-$20d45m37,4s & \small SAB(s)d & \small 3.27 & \small 9.86 & \small $-$19.24 & \small 21.4 & \small 6.9 & \small 71 & \small  & \small  & \small  & \small  & \small x & \small \small x & \small x & \small  & \small x \\ 
\hline  
\small NGC\,337A & \small 01h01m33,90s & \small $-$07d35m17,7s  & \small SAB(s)dm & \small 8.13 & \small 12,7B & \small $-$18.67 & \small 5.9 & \small 4.5 & \small 41 & \small  & \small  & \small  & \small  & \small  & \small x & \small x & \small  & \small  \\ 
\hline  
\small IC\,1613 & \small 01h04m47,79s & \small +02d07m04,0s & \small IB(s)m & \small 0.72 & \small 9.88 & \small $-$14.55 & \small 16.2 & \small 14.5 & \small 27 & \small 7.86 & \small 2 & \small  & \small   & \small  & \small x & \small x & \small  & \small x \\ 
\hline  
\small M33 & \small 01h33m50,89s & \small +30d39m36,8s & \small SA(s)cd & \small 0.87 & \small 6.27 & \small $-$18.94 & \small 70.8 & \small 41.7 & \small 54 & \small 8.48 & \small 1 & \small  & \small  & \small x & \small x & \small x & \small  & \small x \\ 
\hline  
\small NGC\,628 & \small 01h36m41,75s & \small +15d47m01,2s & \small SA(s)c & \small 8.985 & \small 9.95 & \small $-$20.68 & \small 10.5 & \small 9.5 & \small 24 & \small 8.78 & \small 1 & \small  & \small x & \small x & \small x & \small x & \small x & \small x \\ 
\hline  
\small IC\,1727 & \small 01h47m29,89s & \small +27d20m00,1s & \small SB(s)m & \small 6.96 & \small 12.07 & \small $-$18.32 & \small 6.9 & \small 3.1 & \small 63 & \small  & \small  & \small  & \small  & \small  & \small \small x & \small x & \small  & \small x \\ 
\hline  
\small NGC\,672 & \small 01h47m54,52s & \small +27d25m58,0s & \small SB(s)cd & \small 7.32 & \small 11.47 & \small $-$18.94 & \small 7.2 & \small 2.6 & \small 69 & \small  & \small  & \small  & \small  & \small x & \small x & \small x & \small  & \small x \\ 
\hline  
\small NGC\,925 & \small 02h27m16,88s & \small +33d34m45,0s & \small SAB(s)d & \small 7.85 & \small 10.69 & \small $-$20.05 & \small 10.5 & \small 5.9 & \small 56 & \small 8.48 & \small 1 & \small  & \small  & \small x & \small x & \small x & \small x & \small x \\ 
\hline  
\small NGC\,1042 & \small 02h40m23,97s & \small $-$08d26m00,8s & \small SAB(rs)cd & \small 7.85 & \small 11,5B & \small $-$17.51 & \small 4.7 & \small 3.6 & \small 39 & \small  & \small  & \small  & \small  & \small  & \small x & \small x & \small  & \small x \\ 
\hline  
\small M77 & \small 02h42m40,71s & \small $-$00d00m47,8s & \small (R)SA(rs)b & \small 10.58 & \small 9.61 & \small $-$20.58 & \small 7.1 & \small 6 & \small 32 & \small 8.64 & \small 1 & \small  & \small  & \small x & \small x & \small x & \small x & \small x \\ 
\hline  
\small NGC\,1058 & \small 02h43m30,00s & \small +37d20m28,8s & \small SA(rs)c & \small 5.2 & \small 11,2V & \small $-$18.89 & \small 3 & \small 2.8 & \small 24 & \small 8.64 & \small 1 & \small  & \small  & \small  & \small  & \small  & \small  & \small  \\ 
\hline  
\small NGC\,1073 & \small 02h43m40,52s & \small +01d22m34,0s & \small SB(rs)c & \small 7.37 & \small 11.47 & \small $-$19.87 & \small 4.9 & \small 4.5 & \small 24 & \small  & \small  & \small  & \small  & \small  & \small  & \small x & \small  & \small x \\ 
\hline  
\small NGC\,1156 & \small 02h59m42,30s & \small +25d14m16,2s & \small IB(s)m & \small 6.09 & \small 12.32 & \small $-$18.56 & \small 3.3 & \small 2.5 & \small 42 & \small 8.16 & \small 9 & \small  & \small  & \small x & \small x & \small  & \small x & \small x \\ 
\hline  
\small NGC\,2283 & \small 06h45m52,69s & \small $-$18d12m37,2s & \small SB(s)cd & \small 9.92 & \small 12.93 & \small $-$18.93 & \small 3.6 & \small 2.8 & \small 41 & \small  & \small  & \small x & \small  & \small  & \small  & \small  & \small  & \small  \\ 
\hline  
\small NGC\,2903 & \small 09h32m10,11s & \small +21d30m03,0s & \small SAB(rs)bc & \small 7.99 & \small 9.68 & \small $-$21.02 & \small 12.6 & \small 6 & \small 61 & \small 8.82 & \small 1 & \small x & \small  & \small x & \small x & \small x & \small x & \small x \\ 
\hline  
\small Sextans\,B & \small 10h00m00,10s & \small +05d19m56,0s & \small IB(s)m & \small 1.55 & \small 11.85 & \small $-$14.39 & \small 5.1 & \small 3.5 & \small 46 & \small 7.5 & \small 3 & \small  & \small  & \small  & \small x & \small x & \small  & \small x \\ 
\hline  
\small Sextans\,A & \small 10h11m00,80s & \small $-$04d41m34,0s & \small IBm & \small 1.43 & \small 11.86 & \small $-$13.56 & \small 5.9 & \small 4.9 & \small 34 & \small 7.53 & \small 3 & \small  & \small  & \small  & \small x & \small x & \small  & \small x \\ 
\hline  
\small UGC\,5829 & \small 10h42m41,91s & \small +34d26m56,0s & \small Im & \small 8 & \small 13.73 & \small $-$16.73 & \small 4.7 & \small 4.2 & \small 27 & \small  & \small  & \small  & \small  & \small  & \small x & \small x & \small  & \small  \\ 
\hline  
\small NGC\,3344 & \small 10h43m31,15s & \small +24d55m20,0s & \small (R)SAB(r)bc & \small 12.44 & \small 10.45 & \small $-$19.64 & \small 7.1 & \small 6.5 & \small 24 & \small 8.72 & \small 1 & \small  & \small x & \small x & \small x & \small x & \small x & \small x \\ 
\hline  
\small M95 & \small 10h43m57,70s & \small +11d42m13,7s & \small SB(r)b & \small 9.97 & \small 11,4g & \small $-$19.84 & \small 3.07 & \small 2.86 & \small 47 & \small 8.82 & \small 1 & \small x & \small x & \small x & \small x & \small x & \small x & \small x \\ 
\hline  
\small NGC\,3377A & \small 10h47m22,30s & \small +14d04m10,0s & \small SAB(s)m & \small 7.44 & \small 14.22 & \small $-$15.64 & \small 2.2 & \small 2.1 & \small 17 & \small  & \small  & \small  & \small  & \small  & \small  & \small  & \small  & \small  \\ 
\hline  
\small M66 & \small 11h20m14,96s & \small +12d59m29,5s & \small SAB(s)b & \small 9.59 & \small 9.65 & \small $-$21.21 & \small 9.1 & \small 4.2 & \small 63 & \small 8.34 & \small 4 & \small x & \small x & \small x & \small x & \small x & \small x & \small x \\ 
\hline  
\small NGC\,3631 & \small 11h21m02,87s & \small +53d10m10,4s & \small SA(s)c  & \small 10.32 & \small 11.01 & \small $-$21.02 & \small 5 & \small 4.8 & \small 17 & \small 8.71 & \small 1 & \small  & \small  & \small  & \small  & \small x & \small x & \small x \\ 
\hline  
\small NGC\,3642 & \small 11h22m17,90s & \small +59d04m28s & \small SA(r)bc & \small 8.378 & \small 12.6 & \small $-$20.57 & \small 1.76 & \small 1.53 & \small 34 & \small  & \small  & \small  & \small  & \small  & \small  & \small  & \small  & \small x \\ 
\hline  
\small NGC\,4027 & \small 11h59m30,17s & \small $-$19d15m54,8s & \small SB(s)dm & \small 12.24 & \small 11.66 & \small $-$20.66 & \small 3.2 & \small 2.4 & \small 41 & \small  & \small  & \small  & \small  & \small  & \small  & \small x & \small  & \small x \\ 
\hline  
\small NGC\,4151 & \small 12h10m32,58s & \small +39d24m20,6s & \small (R')SAB(rs)ab & \small 9.92 & \small 11.5 & \small $-$17.30 & \small 6.3 & \small 4.5 & \small 45 & \small  & \small  & \small  & \small  & \small x & \small  & \small x & \small x & \small x \\ 
\hline  
\small NGC\,4214 & \small 12h15m39,17s & \small +36d19m36,8s & \small IAB(s)m & \small 2.98 & \small 10.24 & \small $-$17.46 & \small 8.5 & \small 6.6 & \small 39 & \small 8.2 & \small 9 & \small  & \small  & \small x & \small x & \small x & \small x & \small x \\ 
\hline  
\small NGC\,4242 & \small 12h17m30,18s & \small +45d37m09,5s & \small SAB(s)dm & \small 6.4 & \small 11,2B & \small $-$17.46 & \small 5 & \small 3.8 & \small 41 & \small  & \small  & \small  & \small x & \small x & \small x & \small x & \small  & \small x \\ 
\hline  
\small M106 & \small 12h18m57,50s & \small +47d18m14,3s & \small SAB(s)bc & \small 7.28 & \small 8,41V & \small $-$20.94 & \small 18.6 & \small 7.2 & \small 67 & \small 8.54 & \small 1 & \small  & \small x & \small x & \small x & \small x & \small x & \small x \\ 
\hline  
\small NGC\,4314 & \small 12h22m31,82s & \small +29d53m45,2s & \small SB(rs)a & \small 9.7 & \small 11.43 & \small $-$19.90 & \small 4.2 & \small 3.7 & \small 27 & \small  & \small  & \small  & \small  & \small  & \small  & \small  & \small  & \small  \\ 
\hline  
\small NGC\,4395 & \small 12h25m48,86s & \small +33d32m48,9s & \small SA(s)m: & \small 4.23 & \small 10.64 & \small $-$18.51 & \small 13.2 & \small 11 & \small 34 & \small 8.32 & \small 9 & \small  & \small x & \small x & \small x & \small x & \small  & \small x \\ 
\hline  
\small NGC\,4449 & \small 12h28m11,10s & \small +44d05m37,1s & \small IBm & \small 3.86 & \small 9.99 & \small $-$19.17 & \small 6.2 & \small 4.4 & \small 45 & \small 8.26 & \small 9 & \small  & \small x & \small x & \small x & \small x & \small  & \small x \\ 
\hline  
\small UGC\,7608 & \small 12h28m44,20s & \small +43d13m26,9s & \small Im & \small 8.25 & \small 13.67 & \small $-$16.76 & \small 3.4 & \small 3.3 & \small 12 & \small  & \small  & \small  & \small  & \small  & \small x & \small x & \small  & \small  \\ 
\hline  
\small NGC\,4490 & \small 12h30m36,24s & \small +41d38m38,0s & \small SB(s)d pec & \small 6.21 & \small 10.22 & \small $-$21.49 & \small 6.3 & \small 3.1 & \small 61 & \small 8.29 & \small 1 & \small  & \small x & \small x & \small x & \small x & \small x & \small x \\ 
\hline  
\small UGC\,7698 & \small 12h32m54,39s & \small +31d32m28,0s & \small Im & \small 4.21 & \small 13 & \small $-$15.70 & \small 6.5 & \small 4.5 & \small 46 & \small 8.2 & \small 7 & \small  & \small  & \small  & \small x & \small x & \small  & \small x \\ 
\hline  
\small NGC\,4618 & \small 12h41m32,85s & \small +41d09m02,8s & \small SB(rs)m & \small 7.24 & \small 11.22 & \small $-$19.44 & \small 4.2 & \small 3.4 & \small 36 & \small  & \small  & \small  & \small  & \small  & \small x & \small x & \small  & \small x \\ 
\hline  
\end{tabular}
\end{table}
\end{landscape}
\pagebreak

\setcounter{table}{2}
\begin{landscape}
\begin{table}
\caption{\textbf{Continued:} List of targets for SIGNALS.}
\small
\begin{tabular}{|c|c|c|c|c|c|c|c|c|c|c|c|c|c|c|c|c|c|c|}
\hline
\small \makecell{%
  ID \\
  (1)
} & \small \makecell{%
  RA \\
  (2)
} & \small \makecell{%
  DEC \\
  (3)
} & \small \makecell{%
  Morphology \\
  (4)
} & \small \makecell{%
  D \\
  (5)
} & \small \makecell{%
  m \\
  (6)
} & \small \makecell{%
  M \\
  (7)
} & \small \makecell{%
  a \\
  (8)
} & \small \makecell{%
  b \\
  (9)
} & \small \makecell{%
  \textit{i} \\
  (10)
} & \small \makecell{%
  Z \\
  (11)
} & \small \makecell{%
  Ref$\_$Z \\
  (12)
} & \scriptsize \makecell{%
  PHANGS \\
  \small (13)
} & \scriptsize \makecell{%
  LEGUS \\
  \small (14)
} & \scriptsize \makecell{%
  HST \\
  \small (15)
} & \scriptsize \makecell{%
  GALEX \\
  \small (16)
} & \small \makecell{%
  \textit{Spitzer} \\
  \small (17)
} & \scriptsize \makecell{%
  CO \\
  \small (18)
} & \scriptsize \makecell{%
  VLA \\
  \small (19)
} \\
\hline  
\small M94 & \small 12h50m53,06s & \small +41d07m13,6s & \small (R)SA(r)ab & \small 5.11 & \small 8.99 & \small $-$19.67 & \small 11.2 & \small 9.1 & \small 36 & \small 8.57 & \small 1 & \small  & \small  & \small x & \small x & \small   & \small x & \small x \\ 
\hline 
\small IC\,4182 & \small 13h05m49,54s & \small +37d36m17,6s & \small SA(s)m & \small 4.21 & \small 13 & \small $-$15.74 & \small 6 & \small 5.5 & \small 24 & \small  & \small  & \small  & \small  & \small   & \small x & \small x & \small  & \small  \\ 
\hline  
\small M63 & \small 13h15m49,33s & \small +42d01m45,4s & \small SA(rs)bc & \small 7.72 & \small 9.31 & \small $-$20.89 & \small 12.6 & \small 7.2 & \small 55 & \small 8.87 & \small 1 & \small  & \small  & \small x & \small x & \small x & \small x & \small x \\ 
\hline
\small NGC\,5068 & \small 13h18m54.80s & \small $-$21d02m21,0s & \small SAB(rs)cd & \small 5.989 & \small 10.52 & \small $-$18,54 & \small 7.2 & \small 6.3 & \small 29 & \small  & \small  & \small x & \small  & \small  & \small x & \small x & \small  & \small x \\ 
\hline  
\small NGC\,5204 & \small 13h29m36,51s & \small +58d25m07,4s & \small SA(s)m & \small 5.22 & \small 11.73 & \small $-$17.01 & \small 5 & \small 3 & \small 53 & \small  & \small  & \small  & \small  & \small  & \small x & \small x & \small  & \small x \\ 
\hline  
\small M51 & \small 13h29m52,71s & \small +47d11m42,6s & \small SA(s)bc pec & \small 7.18 & \small 8.96 & \small $-$21.33 & \small 11.2 & \small 6.9 & \small 52 & \small 8.88 & \small 1 & \small  & \small x & \small x & \small x & \small x & \small x & \small x \\ 
\hline  
\small NGC\,5247 & \small 13h38m03,04s & \small $-$17d53m02,5s & \small SA(s)bc & \small 9.35 & \small 10,5B & \small $-$21.31 & \small 5.6 & \small 4.9 & \small 29 & \small  & \small  & \small  & \small  & \small x & \small  & \small x & \small x & \small x \\ 
\hline  
\small M101 & \small 14h03m12,54s & \small +54d20m56,2s & \small SAB(rs)cd & \small 6.85 & \small 8.31 & \small $-$20.97 & \small 28.8 & \small 26.9 & \small 21 & \small 8.71 & \small 1 & \small  & \small x & \small x & \small x & \small x & \small x & \small x \\ 
\hline  
\small NGC\,5474 & \small 14h05m01,61s & \small +53d39m44,0s & \small SA(s)cd pec & \small 4.34 & \small 11.28 & \small $-$18.07 & \small 4.8 & \small 4.3 & \small 27 & \small 8.19 & \small 1 & \small  & \small x & \small x & \small x & \small x & \small x & \small x \\ 
\hline  
\small NGC\,5585 & \small 14h19m48,20s & \small +56d43m44,6s & \small SAB(s)d & \small 7.37 & \small 11.2 & \small $-$18.71 & \small 5.8 & \small 3.7 & \small 50 & \small  & \small  & \small  & \small  & \small x & \small x & \small x & \small  & \small x \\ 
\hline  
\small UGC\,10310 & \small 16h16m18,35s & \small +47d02m47,1s & \small SB(s)m & \small 7.91 & \small 13.58 & \small $-$17.41 & \small 2.8 & \small 2.2 & \small 37 & \small  & \small  & \small  & \small  & \small  & \small  & \small x & \small  & \small  \\ 
\hline  
\small NGC\,6814 & \small 19h42m40,64s & \small $-$10d19m24,6s & \small SAB(rs)bc & \small 11.75 & \small 12.06 & \small $-$21.57 & \small 3 & \small 2.8 & \small 21 & \small  & \small  & \small  & \small  & \small  & \small  & \small  & \small  & \small x \\ 
\hline  
\small NGC\,6822 & \small 19h44m57,74s & \small $-$14d48m12,4s & \small IB(s)m & \small 0.52 & \small 9.31 & \small $-$15.02 & \small 15.5 & \small 13.5 & \small 29 & \small 8.06 & \small 8 & \small  & \small  & \small x & \small x & \small x & \small   & \small x \\ 
\hline  
\small NGC\,6946 & \small 20h34m52,30s & \small +60d09m14,0s & \small SAB(rs)cd & \small 5.545 & \small 8.23 & \small $-$20,90 & \small 11.5 & \small 9.8 & \small 32 & \small  & \small  & \small  & \small  & \small x & \small  & \small x & \small x & \small x \\ 
\hline  
\small UGC\,12082 & \small 22h34m10,82s & \small +32d51m37,8s & \small Sm& \small 9.79 & \small 14.1 & \small $-$17.14 & \small 2.6 & \small 2.2 & \small 32 & \small  & \small  & \small  & \small  & \small  & \small  & \small  & \small  & \small  \\ 
\hline  
 \small UGC\,12632 & \small 23h29m58,67s & \small +40d59m24,8s & \small Sm: & \small 9.44 & \small 12.78 & \small $-$17.69 & \small 4.5 & \small 3.7 & \small 34 & \small  & \small  & \small  & \small  & \small  & \small  & \small  & \small  & \small  \\ 
\hline
\end{tabular}
		\begin{tablenotes}
		    \item[(1)] $[1]$ Identification of the galaxy
		    \item[(2)] $[2]$ Right ascension (2000)
		    \item[(3)] $[3]$ Declination (2000)
		    \item[(4)] $[4]$ Morphology from the Third Reference Catalogue of Bright Galaxies (RC3; \citealt{Corwin1994}) 
		    \item[(5)] $[5]$ Mean distance [Mpc] from NED
		    \item[(6)] $[6]$ Apparent magnitude [mag] (note B or V if the only one available)
		    \item[(7)] $[7]$ Absolute magnitude [mag]
		    \item[(8)] $[8]$ Major axis size [arcmin]
		    \item[(9)] $[9]$ Minor axis size [arcmin]
		    \item[(10)] $[10]$ Inclination [deg] from the RC3 
		    \item[(11)] $[11]$ Estimated global metallicity (see item 12)
		    \item[(12)] $[12]$ Reference for the global metallicity estimate
                \item[(11.4)] $\,\,\,\,\,\,\,\,\,\,\,\,1)$ \cite{pilyugin2014}, 2) \cite{Sakai2004}, 3) \cite{Kniazev2005}, 4) \cite{Moustakas2010}, 7) \cite{Hunter1999}, 8) \cite{HernandezM2009}, 
                \item[(11.8)] $\,\,\,\,\,\,\,\,\,\,\,\,9)$ \cite{Pilyugin2015}, 10) \cite{Skillman1989}, 11) \cite{Skillman1997}, 12) \cite{vanZee2006}, 
		    \item[(13)] $[13]$ Target included in the PHANGS survey (x)
		    \item[(14)] $[14]$ Target included in the LEGUS survey (x)
		    \item[(15)] $[15]$ Target with observations available from HST (x)
		    \item[(16)] $[16]$ Target with observations available from GALEX (x)
		    \item[(17)] $[17]$ Target with observations available from \textit{spitzer} (x)
		    \item[(18)] $[18]$ Target with CO observations available (x)
		    \item[(19)] $[19]$ Target with HI observations available from the Very Large Array (x)
		    \end{tablenotes}
		\end{table}
\end{landscape}

\bsp	
\label{lastpage}
\end{document}